\documentstyle[aps,psfig]{revtex}
\newcommand{\PP}{{\rm I\hspace{-0.65ex}P}}
\newcommand{\bra}[1]{\mbox{$\langle #1|$}}
\newcommand{\ket}[1]{\mbox{$|#1\rangle$}}

\begin{document}

\begin{title}
{
\hfill{\small {\bf MKPH-T-01-04}}\\
{\bf Offshell effects  in electromagnetic 
reactions on the deuteron} \footnote[2]
{Supported by the Deutsche Forschungsgemeinschaft (SFB 443).}
}
\end{title}

\author
{Michael Schwamb and Hartmuth Arenh\"ovel}
\address{Institut f\"ur Kernphysik,           
         Johannes Gutenberg-Universit\"at,  
         D-55099 Mainz, Germany }
\maketitle

\begin{abstract}
Offshell contributions to the electromagnetic nuclear current
  are evaluated within a 
nonrelativistic approach  by incorporation one-pion loop 
contributions in time-ordered perturbation theory. By construction, the 
correct experimental onshell properties of the nucleon current are 
ensured
 so that only the genuine offshell effects appear as model 
dependent. For a qualitative assessment of such offshell 
effects, this model is applied to photodisintegration 
 of the deuteron for photon energies up to 
 500 MeV. While at low energies offshell contributions are small,
 above 300 MeV they    lead to sizeable effects in  observables
 up to about 30 percent pointing to the necessity of 
incorporating such effects if one aims at theoretical predictions of 
high precision.
\end{abstract}

\section{Introduction}\label{kap1}
The electromagnetic probe provides one of the most  important and 
accurate tools for the study of nuclear or more general hadron structure. 
The probing operator hereby is the electromagnetic current of the system, 
and thus its knowledge is essential for unraveling the internal dynamics. 
The leading contribution is provided by the one-body currents of the 
constituents supplemented by additional interaction currents, 
in general of many-body type. 

The one-body nucleon current has the well-known onshell form 
\begin{equation}\label{1_1}
\bra{N \vec{p}^{\, \prime}} j^{\mu}(q) \ket{N \vec{p}\,}
 = e \, \bar{u}(p^{\prime}) \Big[F_1(q^2) \gamma^{\mu} 
  + i \,\frac{ F_2(q^2)}{2M_N}\,\sigma^{\mu \nu} q_{\nu} \Big] u(p) \,,
\end{equation}
where $e$ denotes the elementary charge. This form follows from 
general principles, i.e., Lorentz covariance, time, parity and gauge 
invariance, provided the nucleon in the initial and final state is 
on mass shell. The Dirac and Pauli form factors, $F_1(q^2)$ and $F_2(q^2)$, 
respectively, reflect the internal structure of the nucleon. In other words, 
nucleon degrees of freedom (d.o.f.) are not the fundamental d.o.f.\, but 
effective d.o.f. These form factors are determined by the 
nucleon's charge and magnetization densities and can be measured 
experimentally in elastic electron-nucleon scattering, leaving aside the 
problem associated with the absence of free neutron targets. It is 
convenient to exhibit explicitly their isospin dependence 
\begin{equation}\label{1_2}
F_i = \frac{1}{2} (F_i^s + \tau_0 F_i^v) \,\, .
\end{equation}
At the photon point the form factors are normalized according to
 \begin{equation}\label{1_3}
F_1^s(0)\equiv e^s_{phys} = 1\,, \quad  
F_1^v(0)\equiv e^v_{phys} =1\, , \quad  
F_1^s(0) \equiv \kappa^s_{phys} =-0.12\, , \quad  
F_1^v(0) \equiv  \kappa^v_{phys} = 3.70\, ,
 \end{equation}
where $e^s_{phys}$, $e^v_{phys}$, $\kappa^s_{phys}$, and $\kappa^v_{phys}$ 
 denote the isoscalar ($s$)
and isovector ($v$) parts of  charge and anomalous magnetic moment, 
respectively. 

The problem of offshell continuation arises if one incorporates this current 
into reactions which involve an offshell nucleon. For example, in 
electromagnetic pion production on the nucleon, the nucleon pole diagram 
contains an intermediate offshell nucleon. Other examples are provided 
by electromagnetic reactions on nuclei. In this case, a nucleon 
interacting with the photon is bound inside the nuclear environment and 
thus is necessarily offshell. Furthermore, the struck nucleon is exposed to 
final state interaction effects afterwards so that in general the 
nucleon is offshell both in the initial and final states. Therefore, such 
reactions call for a generalization of the onshell current of (\ref{1_1}) 
to the offshell case. 

About forty years ago, it has been shown by Bincer \cite{Bin60}, that the 
corresponding offshell current contains in general twelve instead of only 
two form factors which, moreover, do depend not only on the photon momentum
$q$ but also on the invariant masses $W$ and $W'$ of the initial and 
final nucleon, respectively. Due to the complexity of this offshell current 
and the fact that one needs a theoretical model for it, one usually neglects 
in theoretical calculations such offshell effects.
The crucial question, however, remains  whether this approximation 
is sufficiently accurate with respect to the accuracy of experimental data 
to be compared with. In particular, for polarization observables, which play 
an increasing role in present state-of-the-art nuclear physics, it is well
known that even ``small'' reaction mechanisms like offshell effects may 
become important. 

 In the literature, the problem of offshell form factors in the
 $\gamma N$-vertex has been faced in different ways. 
 They are  based either on dispersion theory~\cite{Nym70,Nym71,Har72}, on
 chiral perturbation theory~\cite{BoK93} or on 
 one-pion loop models~\cite{NaK87,TiT90,BoS92,DoS95}.
 Quantitatively, these approaches yield 
quite different results. For example, as has been stressed in~\cite{KoM98}, 
at the photon point the various predictions for the slope of the offshell 
continuation $F^{++}$ of the onshell Pauli form factor taken at the onshell 
point
\begin{equation}\label{1_4}
\left. \frac{ \partial F^{++}(q^2=0, W,W')}{\partial (W/M_N)} 
\right|_{W=W'=M_N}\,
\end{equation} 
differ in magnitude and even in  sign. Thus at present, these numbers 
should only be considered as an illustration. But a more
 quantitative estimate of 
offshell effects in electromagnetic reactions is still missing. Moreover, one 
should keep in mind that offshell effects are not only model but also 
representation dependent and as such are not observable directly. 
This has been stressed recently by Fearing and Scherer~\cite{Fea00,Sch01}, 
who have proven the impossibility of {\it measuring} off-shell effects 
unambiguously in nucleon-nucleon bremsstrahlung and related processes. 

With respect to the study of offshell effects in electromagnetic reactions 
on  nuclei, we would like to mention first the work of Naus and 
Koch~\cite{NaK87}, who have calculated the $e$-$N$ cross section for a bound 
nucleon making some simplifying assumptions about the wave function 
and the current structure. Within their framework, which did not contain 
any nuclear dynamics, they obtained by offshell effects an increase of
about 8~\%  in the cross section. 
Similarly, Tiemeijer and Tjon~\cite{TiT90} estimated offshell effects to 
be of the order of a few percent only in elastic and inelastic photon 
scattering off a nucleus. On the other hand, Song et al.~\cite{SoC92} obtained 
a reduction of the longitudinal response function $R_L$ for $^{12}$C up to 
about 20~\% by electromagnetic offshell effects using a relativistic 
Fermi-gas model. Last but not least, we would like to mention the recent 
work of Kondratyuk and collaborators \cite{KoM98,KoS99}. In~\cite{KoM98}, 
offshell effects were investigated in proton-proton bremsstrahlung using a 
parameterization of the offshell form factors based on their general 
properties and consistent with the above mentioned model calculations
 \cite{BoS92,DoS95}.  Quite important offshell effects were found, especially 
for the analyzing power, which where of the order of about 25 - 30~\%. 
On the other hand, in~\cite{KoS99} a microscopic model for the half offshell 
$\gamma N$-current has been presented by dressing the $\gamma N$-vertex
with pion loops. By implementing the latter in a $K$-matrix approach
to the Compton scattering amplitude on the nucleon, 
 the authors found a sizeable 
enhancement  of up to about 10~\%  for the forward cross section above
 the $\pi$-threshold.

In view of the fact, that all the approaches to offshell effects
 in electromagnetic reactions on  nuclei lack a certain degree of 
consistency, since the underlying dynamical origin of such offshell effects 
is certainly related to the mechanisms responsible for the $NN$-interaction,
we would like to present in this paper a study of offshell effects in 
electromagnetic reactions on the deuteron, where we try to be as 
consistent as possible with respect to the dynamical input.  Therefore,
we use a three-body approach, described in detail
 in \cite{ScA99,ScA00},  whose basic interactions, apart from
 other contributions not important for the forthcoming discussion, 
 consist of  meson-nucleon vertices, electromagnetic
 baryonic and mesonic one-body currents as well as  the
 vertex- and the Kroll-Rudermann contribution.
 These interactions are the basic 
 building blocks of the corresponding effective operators
  like the retarded NN-interaction and  retarded MEC 
  as well as the hadronic and electromagnetic loop corrections.
 The incorporation of the latter lead  directly to the desired
 offshell contributions to the electromagnetic one-body current
 as will be outlined in the next sections. Due to our framework,
 they contain for example the same $\pi$N-vertex like
 the $\pi$-MEC or the one-pion-exchange NN-potential. 

The choice 
of the deuteron is motivated by the fact that the deuteron is the simplest 
nucleus serving as an ideal laboratory for the study of the relevant 
mechanisms of the strong interaction. A variety of such mechanisms like 
meson exchange currents and isobar configurations have first been studied
for the deuteron. For these reasons, we can expect that offshell effects
in the $\gamma N$-vertex can be studied most reliably for reactions on 
this nucleus. 
Moreover, the question of offshell effects in the deuteron is also relevant 
with respect to its use as a neutron target in order to determine neutron 
properties like, e.g., its electric form factor. Since however, the neutron 
bound in the deuteron is not onshell, one has to assess, among other things,
the size of possible offshell modifications. Another  topic of current  
interest in this respect is the proposed measurement (at MAMI in Mainz and 
ELSA in Bonn) of the spin asymmetry and the GDH-sum rule of the deuteron 
in order to test our knowledge of the spin asymmetry of the neutron.

In order to simplify the discussion, we will consider in this work
only real photons. 
 Moreover, we have selected for this study as basic process
 the simplest photonuclear reaction on a nucleus, i.e.\ 
 deuteron photodisintegration.  The study of more complicated
 processes like electrodisintegration or pionproduction
 is devoted to forthcoming papers.

As has been mentioned above,  we take as theoretical framework
 a previously developed 
nonrelativistic time ordered perturbation approach based on  a model with 
explicit meson, nucleon, and isobar d.o.f.~\cite{ScA99,ScA00}. Similar 
to~\cite{{NaK87},{TiT90}}, we calculate the offshell effects in the 
nucleon one-body current by evaluating a one-pion loop contribution. 
We are aware of the fact that the size of offshell effects will be model
 dependent.  However, at present we 
consider the adopted treatment realistic enough in order to find at least a 
 semiquantitative answer to the question, whether offshell effects in 
electromagnetic reactions on the deuteron are in general negligible or have 
to be included. If the latter turns out to be true, one certainly has to 
refine the present model in the future towards a more quantitative treatment 
of such effects. Comparing the general structure of our model with the 
approach in~\cite{{NaK87},{TiT90}}, we would like to point out at least two
major differences. Whereas in ~\cite{{NaK87},{TiT90}} a fully relativistic
framework has been adopted, we use a nonrelativistic model.
 It allows us to study offshell effects {\it consistently} within a realistic 
interaction model which is suitable for energies not only below, but also 
above $\pi$-threshold, where the $\Delta$ isobar plays an important role.
Moreover, due to a specific renormalization scheme, which will be described 
in detail below, we guarantee that the experimental onshell properties 
of the nucleon current are preserved in our model.  Only the offshell 
continuation of the current is model dependent. As we will see in detail, 
this treatment allows us to preserve the good description of deuteron 
photodisintegration of~\cite{ScA00}. Such a scheme has not been applied 
in~\cite{{NaK87},{TiT90}}, where the offshell as well as the onshell
form factors have been calculated within the same framework. Due to the 
simplicity of the one-pion loop model, the onshell values did not reproduce
the experimental ones. For example, Naus and Koch~\cite{NaK87} obtained 
$\kappa^s_{phys} = -3.19$ and $\kappa^v_{phys} = 4.21$, whereas the
  vector dominance 
model adopted by Tiemeijer and Tjon~\cite{TiT90} yielded $\kappa^s_{phys} 
 = -0.34$ 
and $\kappa^v_{phys}=3.82$ for pseudovector $\pi N$-coupling which are 
 closer to experiment.

In Sects.~\ref{kap2_1} and \ref{kap2_2}, we briefly describe the general 
framework of~\cite{ScA99,ScA00} which we use in the present study. It allows 
to incorporate completely retardation effects in potentials and meson 
exchange currents, and we consider it suitable for the description of 
photoreactions in the energy range up to about $k_{lab}=500$ MeV.
The model for evaluating offshell effects in the electromagnetic one-body
current is described in Sect.~\ref{kap3}. In Sects.~\ref{kap4} and \ref{kap5},
we discuss the predictions for a variety of observables and compare them 
to experimental data where available. In the last section~\ref{kap6}, 
we will draw some conclusions and give an outlook for future studies.

\section{Review of the theoretical framework}\label{kap2_1}

The model Hilbert space consists of three orthogonal subspaces
\begin{equation}\label{kap3_hilbert}
{\cal H}^{[2]} = {\cal H}^{[2]}_{\bar N} \oplus
{\cal H}^{[2]}_{\Delta} \oplus 
{\cal H}^{[2]}_{X}\,,
\end{equation}
where ${\cal H}^{[2]}_{\bar N}$ contains two bare nucleons, 
${\cal H}^{[2]}_{\Delta}$ one nucleon and one $\Delta(1232)$-resonance, 
and ${\cal H}^{[2]}_{X}$ two nucleons and one meson 
$X \in \{ \pi, \rho, \sigma, \delta, \omega, \eta\}$. Thus the latter 
can be divided further into subspaces corresponding to the different mesons
\begin{equation}
{\cal H}^{[2]}_{X}=\sum_{i \in \{\pi, \rho, \sigma, \delta, \omega, \eta\}}
{\cal H}^{[2]}_{X_i}\,.
\end{equation}
In order to distinguish the various sectors of ${\cal H}^{[2]}$, we introduce 
appropriate projection operators by
\begin{eqnarray}\label{kap3_projektoren_delta}
& & P_{{\bar N}}  {\cal H}^{[2]}  =
{\cal H}^{[2]}_{\bar N}\,, \qquad 
P_{\Delta}{\cal H}^{[2]}
={\cal H}^{[2]}_{\Delta}\,,
\qquad  P_{X_i}{\cal H}^{[2]} 
={\cal H}^{[2]}_{X_i}\, . \label{kap3_projektoren_delta2}
\end{eqnarray}
For convenience, we furthermore introduce 
\begin{equation}
 P=P_{\bar N}+P_{\Delta} \qquad \mbox{and} \qquad  
P_X = \sum_{i \in \{\pi, \rho, \sigma, \delta, \omega, \eta\}} P_{X_i} \,. 
\label{kap3_projektoren_delt2} 
\end{equation}
Using the notation 
\begin{equation}\label{matrix_kurz}
\Omega_{\alpha \beta} = P_{\alpha} \Omega P_{\beta}\, ,
\qquad \alpha,\,\beta \in \{\bar N,\, \Delta,\, X \}\,,
\end{equation}
any operator $\Omega$ acting in  ${\cal H}^{[2]}$ can be written as a 
symbolic $3\times 3$ matrix 
\begin{equation}\label{matrixnotation}
\Omega = \left( \begin{array}{ccc}
\Omega_{{\bar N}{\bar N}} & 
\Omega_{{\bar N}\Delta} & 
\Omega_{{\bar N} X} \\
\Omega_{\Delta {\bar N}} & 
\Omega_{\Delta \Delta} & 
\Omega_{\Delta X} \\
\Omega_{X {\bar N}} & \Omega_{X \Delta} & \Omega_{X X} 
                \end{array} \right) .
\end{equation}

\begin{figure}[hp]
\centerline{\psfig{figure=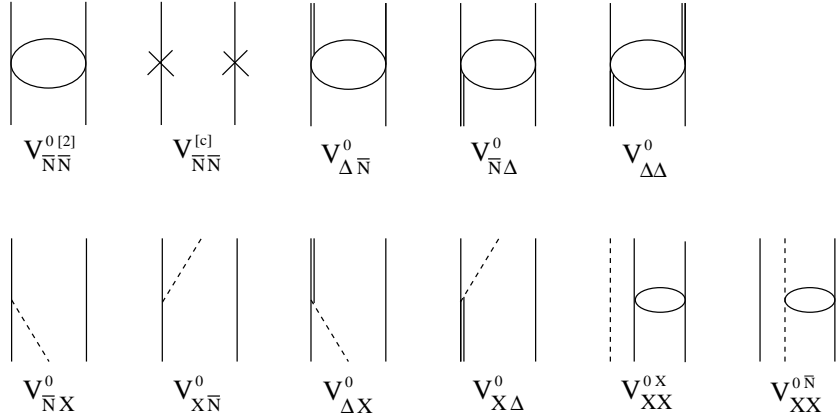,width=11cm,angle=0}}
\vspace{0.5cm}
\caption{Diagrammatic representation of the various components of 
 $V^0$. The open ellipse symbolizes a given hermitean two-body interaction.
 The one-nucleon counter term $v^{[c]}$ is indicated by a cross.}
\label{potentialuebersicht}
\end{figure}

The hamiltonian $H$ is subdivided into a kinetic part $H_0$, containing 
the physical mass of the nucleon, and into an interaction part $V^0$.
As outlined in detail in~\cite{ScA99}, $V^0$ comprises besides baryon-meson 
vertices a two-body interaction $V_{PP}^{0 \, [2]}$, characterized by a 
superscript ``$[2]$'' and represented diagrammatically in 
Fig.~\ref{potentialuebersicht}, and furthermore a counter term  $V^{[c]}$ 
taking into account the nonvanishing mass difference between a bare and a 
physical nucleon. Any hadronic state $\ket{\bar\alpha}$ can be written as 
a sum of a ``baryonic'' part $P\ket{\bar\alpha}$ and a ``mesonic'' part 
$P_X \ket{\bar\alpha}$. For an eigenstate $\ket{\bar\alpha_E}$ of the 
Hamilton operator 
\begin{equation}\label{2_1}
H \ket{\bar\alpha_E}  = E \ket{\bar\alpha_E} \, ,
\end{equation}
the two components are related by 
\begin{equation}\label{2_2}
P_X \ket{\bar\alpha_E} = G^X(E + i \epsilon) V^0_{X P} \ket{\bar\alpha_E}\,,
\end{equation} 
where $G^X(z)$ describes the propagation of two interacting nucleons
in the presence of a spectator meson  
\begin{equation}\label{g_x}
 G^X(z)   =  (z-H_{0,\,XX}-V^0_{XX})^{-1} \,.
 \label{kap3_vqq3_mod}
\end{equation}
The baryonic component $P\ket{\bar\alpha_E}$ is obtained as solution of 
a Schr\"odinger like equation
\begin{equation}\label{2_schr1}
\left(H_0 + V^{0}_{PP} + V^0_{PX} G^X(z) V^0_{XP}  \right)
  P \,\ket{\bar\alpha_E}
= E \,P\,\ket{\bar\alpha_E} \, .
\end{equation}
  Note that   $V^{0}_{PP}$ as well as   
 $V^0_{PX} G^X(z) V^0_{XP} $ consist of a connected ($con$) and 
 a disconnected ($dis$) part,
\begin{eqnarray}
V^{0}_{PP} &=& V^{0,\,dis}_{PP} + V^{0,\,con}_{PP} \,, 
 \quad V^{0,\,dis}_{PP} \equiv V^{[c]}_{PP}\,\, , \quad
V^{0,\,con}_{PP} \equiv V^{0\,[2]}_{PP}\,\, , 
\label{2_erl1}\\
 V^0_{PX} G^X(z) V^0_{XP} &=&
 \left[V^0_{PX} G^X(z) V^0_{XP}\right]_{dis} + 
 \left[ V^0_{PX} G^X(z) V^0_{XP} \right]_{con} \,. \label{2_erl2}
\end{eqnarray}
As has been discussed in~\cite{ScA99}, the disconnected pieces contain by 
definition in general only those parts of the potential which do not describe 
an interaction between the two baryons. The only exception is the $\pi N$ loop 
contribution to the $\Delta$ self energy, which we have incorporated into 
the connected part, containing otherwise the genuine baryon-baryon 
interactions. 

Now, renormalized baryon-baryon and meson-baryon interactions are introduced by
 (consider for example \cite{ScP80,ElF88})
\begin{eqnarray}
 V^{[2]}_{PP}&=&({\widehat Z}_{[2]}^{os})^{-1}\,V_{PP}^{0\,[2]} \,
({\widehat Z}_{[2]}^{os})^{-1}\,, \label{renorm1}\\
V_{XP}&=&({\widehat Z}_{[2]}^{os})^{-1}\,V_{XP}^0\,
({\widehat Z}_{[2]}^{os})^{-1}\,  ,  \label{renorm2}
\end{eqnarray}
in order to arrive at an effective description in terms of the renormalized
 baryonic component 
\begin{equation}\label{2_bar_alpha}
 \ket{\alpha_E} ={\widehat Z}_{[2]}(E+i\epsilon) \, P \ket{\bar{\alpha}_E} \,.
\end{equation} 
Here, the renormalization operator ${\widehat Z}_{[2]}$  and its onshell 
 value ${\widehat Z}^{os}_{[2]}$ are defined by  
\begin{eqnarray}
{\widehat Z}^2_{[2]}(z) &=&  1 + \int dz'\,\delta(z'-H_0)\,
\left[V^0_{{\bar N}X}\,G_0(z')\,G_0(z)\,V^0_{X{\bar N}}
\right]_{dis}\,\, , \label{kap3_renorm_mat_z2_2} \\
({\widehat Z}^{os}_{[2]})^2  &=& \int dz\,\delta(z-H_0)\,
{\widehat Z}^2_{[2]}(z)  \label{2_zos_z} \,\, , 
\end{eqnarray}
 with
\begin{equation}\label{2_g0}
G_0(z) = \left(z-H_0\right)^{-1} \, .
\end{equation}

 We have shown in~\cite{ScA99}, that ${\widehat Z}^{os}_{[2]}$ can be treated
within a very good approximation as an energy- and momentum independent 
constant in ${\cal H}^{[2]}_{\bar N}$. 
The  renormalized interactions (\ref{renorm1}) and (\ref{renorm2})
 contain renormalized coupling constants 
  which are the relevant physical quantities determined either by experiment
 or fit to the experimental data. Furthermore, we introduce an
  auxiliary quantity, namely a ``dressing operator''
\begin{equation}\label{kap3_renorm_mat_r1}
{\widehat R}(z) = {\widehat Z}_{[2]}^{os}\,
{\widehat Z}_{[2]}^{-1}(z)\, , 
\end{equation}
which is equal to the identity for onshell particles.
 It  is diagonal with respect to 
two-body plane waves 
\begin{equation}\label{5_r1}
 \bra{{\bar N} (\vec{p}_1^{\,\prime}\,)\,\, {\bar N} (\vec{p}_2^{\,\prime}\,)} 
{\widehat R}(z) 
 \ket{ \bar N (\vec{p}_1\,)\,\bar N (\vec{p}_2\,)}  \nonumber\\
 = \delta(\vec{p}_1^{\,\prime}-\vec{p}_1) \,\,
 \delta(\vec{p}_2^{\,\prime}-\vec{p}_2) R(z,\vec{p}_1, \vec{p}_2) \,\, ,
\end{equation}
with 
\begin{equation}\label{4_rr2}
R(z,\vec{p}_1, \vec{p}_2) =
\Big[1- \Big(z- e_N(p_1) - e_N(p_2)\Big)
\Big( \Gamma_{\vec{p}_1}(z-e_N(p_2))+ \Gamma_{\vec{p}_2}(z-e_N(p_1))\Big)
\Big]^{-\frac{1}{2}} \,\, , 
\end{equation}
where $e_N(p) = \sqrt{M_N^2 + p^2}$ denotes the nucleon onshell energy 
and $\Gamma_{\vec{p}}$ is defined by
 \begin{eqnarray}\label{4_ss}
 \delta(\vec{p}^{\,\prime}-\vec{p}\,) \,
\Gamma_{\vec{p}}(z) &=& \bra{\bar N (\vec{p}^{\,\prime} \,)} v_{P X_{\pi}} 
( e_N(p) - h_0 )^{-2} \,(z - h_0)^{-1} v_{X_{\pi} P} 
\ket{\bar N (\vec{p}\,)} \,.
\end{eqnarray}
Here we have exploited the fact that the $\pi N$ interaction 
$V_{P X_{\pi}}$ and the free Hamiltonian $H_0$ can be split into corresponding 
one-body parts $v_{P X_{\pi}}$ and $h_0$, acting in the one-nucleon 
sector~\cite{ScA99}.

As has been discussed in detail in \cite{ScA99}, one finally  obtains from 
(\ref{2_schr1}) an equation for $\ket{{\alpha}_E}$ 
\begin{equation}\label{2_schr2}
H_{eff} \ket{{\alpha}_E} = E \ket{{\alpha}_E}
\end{equation}
 where 
\begin{equation}\label{2_heff}
 H_{eff} := 
 H_0 +  {\widehat R}(E+i\epsilon) \Big(V^{[2]}_{PP} +
[V_{P X} G^X(E+i\epsilon) V_{XP}]_{con} \Big){\widehat R}(E+i\epsilon) 
\end{equation}
contains renormalized quantities only. For scattering states, 
$\ket{{\alpha}_E}$ obeys a corresponding Lippmann-Schwinger 
equation~\cite{ScA99}. In our 
explicit calculations, we only solve equations like~(\ref{2_schr2}) which 
incorporate physical couplings.

\section{The electromagnetic current}\label{kap2_2}

\begin{figure}[htp]
\centerline{\psfig{figure=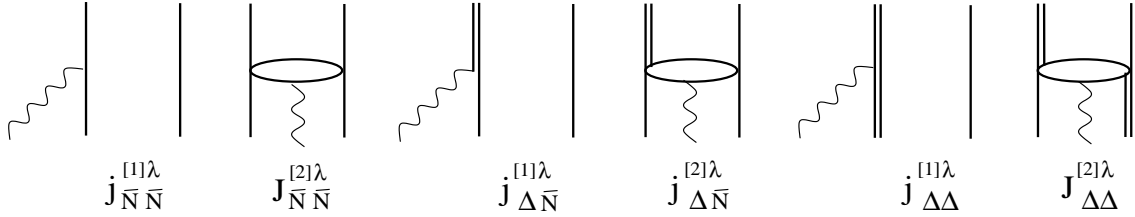,width=15cm,angle=0}}
\vspace{0.5cm}
\caption{Diagrammatic representation  of the baryonic currents. An open  
ellipse symbolizes a two-body exchange current, generated for example by
 heavy meson exchange.}
\label{figem1}
\end{figure}

\begin{figure}[htp]
\centerline{\psfig{figure=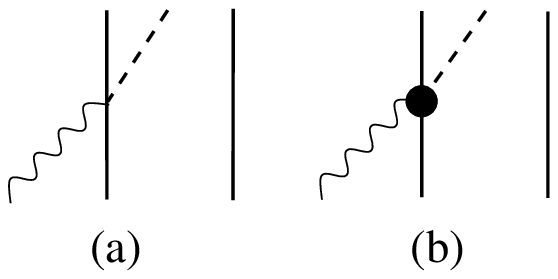,width=6cm,angle=0}}
\vspace{0.5cm}
\caption{Diagrammatic representation of the meson production currents 
$J^{\mu}_{X{\bar N}}$: (a) contact current $j_{X\bar N}^{(1)\,\mu}$ and 
(b) vertex current $j_{X\bar N}^{(1v)\,\mu}$.}
\label{figem2}
\end{figure}

\begin{figure}[htp]
\centerline{\psfig{figure=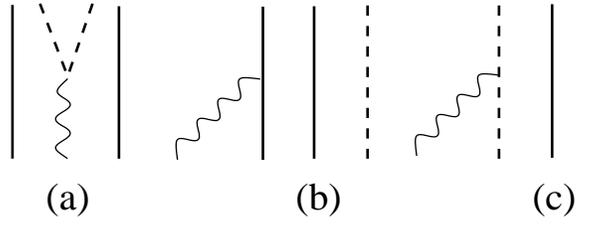,width=8cm,angle=0}}
\vspace{0.5cm}
\caption{Diagrammatic representation of (a) the two-meson production current 
$j^{(0)\, \mu}_{X{\bar N} }$, 
 and the current components 
$J^{\mu}_{XX}$: (b) nucleon current $j^{{\bar N}\, \mu}_{XX}(\vec{k})$,
  (c) meson current   $j^{X \, \mu}_{XX}(\vec{k})$.}
\label{figem3}
\end{figure}

In this section, we briefly review the general structure of the
electromagnetic current, which has been discussed in detail in~\cite{ScA00}. 
We distinguish between three types of currents which are graphically 
depicted in Figs.~\ref{figem1} through  \ref{figem3}. The first type 
comprises the one- and two-body pure baryon currents, labeled by  
superscripts ``[1]'' and ``[2]'', respectively. As outlined in~\cite{ScA00}, 
the two-body parts describe among other things 
 effective heavy meson exchange currents, which 
are not generated explicitly via the elementary meson-baryon vertices and 
the meson production/annihilation currents. The second current type, 
represented in Fig.~\ref{figem2}, comprises the ``one-meson 
production/annihilation'' currents consisting of the Kroll-Rudermann
contact term $j^{(0)\, \mu}_{X {\bar N}}$ and a vertex current
$j^{(1)\, \mu}_{X {\bar N}}$, where the lower case  letter $j$ from here on 
always denotes a one-body-current. 
The third current type is shown in  Fig.~\ref{figem3} and
 describes the coupling of a 
photon to a meson or nucleon in the presence of two spectator nucleons or 
one nucleon and one meson, respectively, and the annihilation/creation of 
two mesons by a photon. One should note that the inclusion of currents 
depicted in Figs.~\ref{figem2} and \ref{figem3} allows us to incorporate 
full $\pi$-retardation in the two-body meson-exchange currents.

In view of the effective description in terms of baryonic d.o.f.\ only, 
it is useful to introduce an ``effective'' current operator, acting in 
pure baryonic space, by
\begin{equation}\label{kap5_effektiv_def}
 \bra{{\alpha}_{E'}}  J_{eff}^{\mu}(z,\vec{k})  \ket{{\alpha}_E}
  := \bra{\bar{\alpha}_{E'}} J^{\mu}(\vec{k}) \ket{\bar{\alpha}_E} \,,
\end{equation}  
where $\vec{k}$ denotes the photon momentum in the c.m.\ frame.
Compared to \cite{ScA00}, we have changed our definition of the
 effective current operator for reasons of convenience.
   Denoting by 
 ${\widetilde J}_{eff}^{\mu}(z,\vec{k})$ the effective current of \cite{ScA00},
 one has the relation 
\begin{equation}\label{change}
  J_{eff}^{\mu}(z,\vec{k}) =
  \frac{{\widehat R}(z)}{{\widehat Z}^{os}_{[2]}} \,\, 
 {\widetilde J}_{eff}^{\mu}(z,\vec{k}) 
   \frac{{\widehat R}(z-k)}{{\widehat Z}^{os}_{[2]}} \,\,
\end{equation}
 to     $J_{eff}^{\mu}(z,\vec{k})$ of  (\ref{kap5_effektiv_def}).
   Obviously, this change of notation is  irrelevant for our numerical results.

\begin{figure}[tp]
\centerline{\psfig{figure=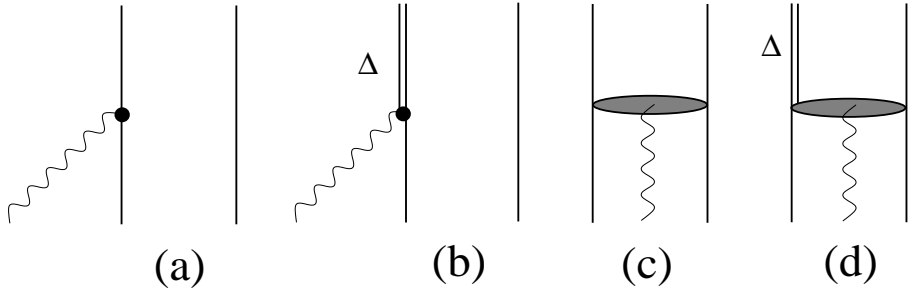,width=12cm,angle=0}}
\vspace{0.5cm}
\caption{Diagrammatic representation of the separate contributions
to the effective current operator $J_{eff}^{\mu}(z,\vec{k})$ (see
Eq.~(\ref{kap5_eff_zerlegung})):
 (a) $J_{eff}^{N[1]\, \mu}(z,\vec{k})$,
 (b) $J_{eff}^{\Delta[1]\, \mu}(z,\vec{k})$,
 (c) $J_{eff}^{N[2]\, \mu}(z,\vec{k})$, and
 (d) $J_{eff}^{\Delta[2]\, \mu}(z,\vec{k})$.}
\label{figem11}
\end{figure}

The effective current operator can be split into a nucleonic and a resonant 
part with superscripts ``$N$'' and ``$\Delta$'', respectively, which in turn 
are divided into one- and two-body pieces,
\begin{equation}\label{kap5_eff_zerlegung}
 J_{eff}^{\mu}(z,\vec{k}) =
 J_{eff}^{N[1]\, \mu}(z,\vec{k}) +
 J_{eff}^{\Delta[1]\, \mu}(z,\vec{k}) +
 J_{eff}^{N[2]\, \mu}(z,\vec{k}) +
 J_{eff}^{\Delta[2]\, \mu}(z,\vec{k}) \,.
\end{equation}
Their diagrammatic representation is shown in Fig.~\ref{figem11}. Except for 
the effective one-body current  $J_{eff}^{N[1]\, \mu}(z,\vec{k})$,
 we adopt for all 
other currents the explicit expressions given in~\cite{ScA00} (apart from
 the abovementioned change (\ref{change}) in notation), 
which therefore are not repeated here. In this context, one should note 
 that  offshell contributions have already been
 implemented in  the one-body $\gamma {\bar N} \Delta$-current 
 $J_{eff}^{\Delta[1]\, \mu}(z,\vec{k})$ as has been outlined in \cite{ScA00}.
 
The evaluation of the effective nucleonic one-body contribution
 $J_{eff}^{N[1]\, \mu}(z,\vec{k})$ is the basic aim of this paper. It will 
be outlined in the next section.

\section{Offshell contributions to the
 nucleon one-body current}\label{kap3}

By a straightforward calculation, one obtains for the matrix element of the
effective nucleonic one-body current operator $J_{eff}^{N[1]\, \mu}(z,\vec{k})$
between two  eigenstates $\ket{{\alpha}_{W}}$ and $\ket{{\alpha}_{W-k}}$
 of $H_{eff}$ (\ref{2_schr2})  with
invariant energies $W$ and $W-k$, respectively, 
the following expression with $z = W + i \epsilon$ 
\begin{equation}\label{3_eff1}
\bra{{\alpha}_{W}}   J_{eff}^{N[1]\, \mu}(z,\vec{k}) \ket{{\alpha}_{W-k}} 
  = \sum_{i=1}^9 
\bra{\alpha_{W}} 
  \frac{{\widehat R}(z)}{{\widehat Z}^{os}_{[2]}} 
  J_{i}^{N[1]\, \mu}(z,\vec{k})
  \frac{{\widehat R}(z-k)}{{\widehat Z}^{os}_{[2]}} 
  \ket{\alpha_{W-k}} \,\, , 
\end{equation}
 where the various terms, represented diagrammatically in Fig.~\ref{figem5},
 are given by
\begin{eqnarray}
  J_{1}^{N[1]\, \mu}(z,\vec{k}) &=& 
\sum_{i=1,2} j^{[1]\,\mu}_{{\bar N} {\bar N}}(\vec{k},i) \, ,
\label{kap3_i1}\\
  J_{2}^{N[1]\, \mu}(z,\vec{k})
&=& \sum_{i=1,2}
  v^0_{{\bar N}X_{\pi}}(i) G_0(z)
  \, j^{{\bar N}\, \mu}_{X_{\pi} X_{\pi}}(\vec{k}, i) 
 G_0(z-k) v^0_{X_{\pi} {\bar N}}(i)   \, ,
 \label{kap3_i2}\\
  J_{3}^{N[1]\, \mu}(z,\vec{k})
&=& \sum_{i=1,2}
    v^0_{{\bar N}X_{\pi}}(i) G_0(z)
  \, j^{\pi\, \mu}_{X_{\pi} X_{\pi}} (\vec{k})
  G_0(z-k) v^0_{X_{\pi} {\bar N}}(i) \, , 
 \label{kap3_i3} \, \\ 
  J_{4}^{N[1]\, \mu}(z,\vec{k})
&=& \sum_{i=1,2}
   \, j^{(0)\, \mu}_{{\bar N} X_{\pi \pi}} (\vec{k})\,
 G_0(z-k)
  v^0_{X_{\pi \pi}X_{\pi}}(i) G_0(z-k) v^0_{X_{\pi} {\bar N}}(i)\, ,
 \label{kap3_i4}\\
  J_{5}^{N[1]\, \mu}(z,\vec{k})
&=& \sum_{i=1,2}
   v^0_{{\bar N} X_{\pi}}(i) G_0(z) v_{X_{\pi} X_{\pi \pi}}(i)  G_0(z)
  \, j^{(0)\, \mu}_{X_{\pi \pi} {\bar N} } (\vec{k}) \, ,
 \label{kap3_i5}\\
  J_{6}^{N[1]\, \mu}(z,\vec{k})
&=& \sum_{i=1,2}
   \, j^{(1)\, \mu}_{{\bar N} X_{\pi}} (\vec{k},i)\,
 G_0(z-k) v^0_{X_{\pi} {\bar N}}(i)  \, , 
 \label{kap3_i6}\\
  J_{7}^{N[1]\, \mu}(z,\vec{k})
 &=& \sum_{i=1,2 }
  v^{0}_{{\bar N}X_{\pi}}(i) G_0(z)
  \, j^{(1)\, \mu}_{X_{\pi} {\bar N}}(\vec{k},i)\, ,
 \label{kap3_i7}\\
  J_{8}^{N[1]\, \mu}(z,\vec{k}) &=& \sum_{i=1,2}
   \, j^{(1v)\, \mu}_{{\bar N} X_{\pi}} (\vec{k},i)\,
 G_0(z-k) v^{0}_{X_{\pi} {\bar N}}(i)  \, , 
 \label{kap3_i8}\\
  J_{9}^{N[1]\, \mu}(z,\vec{k}) 
&=& \sum_{i=1,2}
  v^{0}_{{\bar N}X_{\pi}}(i) G_0(z)
  \, j^{(1v)\, \mu}_{X_{\pi} {\bar N}}(\vec{k},i)\, .
 \label{kap3_i9} 
\end{eqnarray}

With respect to these relations, a few remarks are in order. Whereas 
$J_{1}^{N[1]\, \mu}(z,\vec{k})$ is identical to the bare 
$\gamma {\bar N}$-current, supplemented by a suitable  
``counter''-current which will be defined below, the remaining terms
describe meson cloud contributions (see Fig.~\ref{figem5}).  In this
work, we restrict ourselves solely to the pion as explicit d.o.f.\
expecting that heavier mesons, due to their much larger mass, will not
change our results and conclusions significantly in the energy region 
considered here. The contributions $J_{4}^{N[1]\, \mu}(z,\vec{k})$ and 
$J_{5}^{N[1]\, \mu}(z,\vec{k})$ contain intermediate two-pion states, which
therefore are violating formally our Hilbert space. However, similar to the 
retarded meson exchange currents discussed in~\cite{ScA00}, these 
contributions have to be taken into account in order to guarantee -- 
at least approximately -- gauge 
invariance as is discussed below. All ingredients (vertices, currents) of
the meson cloud contributions (\ref{kap3_i3}) through (\ref{kap3_i9})
are fixed by the corresponding expressions given in~\cite{ScA99,ScA00}. 
Note that the pion-nucleon vertex $v^{0}_{X_{\pi} {\bar N}}$ contains
a dipole form factor with a cutoff mass of 1700 MeV so that the loop integrals
do not diverge. 

The occurrence of the onshell renormalization operator 
 ${\widehat Z}^{os}_{[2]}$ in the matrix element 
(\ref{3_eff1}) suggests to introduce  ``renormalized'' 
 currents  by defining 
(note the similarity to (\ref{renorm1}) and (\ref{renorm2}))
\begin{equation}\label{3_re}
 J_{i,\,ren}^{N[1]\, \mu}(z,\vec{k}) =
 ({\widehat Z}^{os}_{[2]} )^{-1}
 J_{i}^{N[1]\, \mu}(z,\vec{k}) \,
  ({\widehat Z}^{os}_{[2]} )^{-1} \,\, 
 \quad  i \in \{1,...,9\}\,,
\end{equation}
so that   the corresponding renormalized meson cloud contributions
($i=2$ through 9 in Fig.\ \ref{figem5}) contain solely the physical 
and not the bare pion-nucleon coupling constant. The explicit evaluation 
of the loops can be performed by using standard techniques 
(see Appendix A for further details). For the sake of simplicity, 
we use the nonrelativistic expressions for the pion-nucleon vertex 
$v^{0\, nonrel}_{X_{\pi} {\bar N}}$ and for the momentum-energy relation 
of the intermediate nucleon which absorbs/emits a pion.

Introducing as a shorthand 
\begin{equation}\label{kap3_abb} 
 {\cal J}_{loop}^{\mu}(z,\vec{k}) 
=  \sum_{i=2}^9   J_{i,\,ren}^{N[1]\, \mu}(z,\vec{k})
\end{equation}
for the renormalized loop contributions, one obtains in general for the 
matrix element of the latter between two-body momentum eigenstates 
\begin{eqnarray}
 \bra{{\bar N} (\vec{p}_1^{\,\prime}\,)\,\, {\bar N} (\vec{p}_2^{\,\prime}\,)} 
 & & {\rho}_{loop}(z,\vec{k})
 \ket{ \bar N (\vec{p}_1\,)\,\,\bar N (\vec{p}_2\,)} 
= e \, \delta(\vec{p}_1^{\,\prime}-\vec{p}_1-\vec{k}) \,
 \delta(\vec{p}_2^{\,\prime}-\vec{p}_2) \, 
\nonumber \\
& & 
\times \sum_{a=0,1} \sum_{b,c}
  F^{a,b,c}(z_{sub}(1),p_1^{\prime},k) 
 \left[ \sigma^{[a]}(1) \times \left[ Y^{[b]}(\hat{p_1^{\prime}})
 \times Y^{[c]}(\hat{k}) \right]^{[a]} \right]^{[0]} 
+(1\leftrightarrow 2)\,,\nonumber\\
 \bra{{\bar N} (\vec{p}_1^{\,\prime}\,)\,\, {\bar N} (\vec{p}_2^{\,\prime}\,)} 
 &&   {\vec{\cal J}}_{loop}(z,\vec{k})
 \ket{ \bar N (\vec{p}_1\,)\,\,\bar N (\vec{p}_2\,)} 
= e \, \delta(\vec{p}_1^{\,\prime}-\vec{p}_1-\vec{k}) \,
 \delta(\vec{p}_2^{\,\prime}-\vec{p}_2) \, 
 \nonumber \\
& & 
\times \sum_{a=0,1} \sum_{b,c,d}  G^{a,b,c,d}(z_{sub}(1),p_1^{\prime},k)
 \left[ \sigma^{[a]}(1) \times \left[ Y^{[b]}(\hat{p_1^{\prime}})
 \times Y^{[c]}(\hat{k}) \right]^{[d]} \right]^{[1]} 
+(1\leftrightarrow 2)\,,
\label{kap3_all2}
\end{eqnarray}
where we use the convention $\sigma^{[0]} \equiv 1$,  $k = |\vec{k}|$ and 
 $ p^{\prime} = |\vec{p}^{\,\prime}|$.
The quantity $z_{sub}(1)$ denotes the offshell-energy of the active nucleon 
 "1" on which the photon is absorbed. In time ordered perturbation theory, 
$z_{sub}(1)$ is given by $z_{sub}(1) =z- \sqrt{M_N^2 + \vec{p}_2^{\,2}}$, where
$\vec{p}_2$ is the momentum of the spectator nucleon "2".
  The tensor ranks $b,c$,
and $d$ in (\ref{kap3_all2}) are not fixed unambiguously by the underlying 
fundamental symmetries of the current. In our explicit evaluations, we 
restrict ourselves to the lowest possible ranks,
 i.e.\
\begin{eqnarray}
 \bra{{\bar N} (\vec{p}_1^{\,\prime}\,)\, {\bar N} (\vec{p}_2^{\,\prime}\,)} 
 &&  {\rho}_{loop}(z,\vec{k})
 \ket{ \bar N (\vec{p}_1\,)\,\,\bar N (\vec{p}_2\,)} =   
  e \, \delta(\vec{p}_1^{\,\prime}-\vec{p}_1-\vec{k}) \,
 \delta(\vec{p}_2^{\,\prime}-\vec{p}_2) \, 
 \hat{\alpha}(z_{sub}(1),p_1^{\prime},k) +(1\leftrightarrow 2)
\, ,
\label{kap3_spez1} \\
 \bra{{\bar N} (\vec{p}_1^{\,\prime}\,)\, {\bar N} (\vec{p}_2^{\,\prime}\,)} 
 && \vec{\cal J}_{loop}(z,\vec{k}) 
 \ket{ \bar N (\vec{p}_1\,)\,\,\bar N (\vec{p}_2\,)} =
  e \, \delta(\vec{p}_1^{\,\prime}-\vec{p}_1-\vec{k}) \,
 \delta(\vec p_2^{\,\prime}-\vec{p}_2) \,
\Big(e \hat{\beta}(z_{sub}(1),p_1',k) 2 \vec{p}_1^{\,\prime}  \nonumber \\
&&+  e  \hat{\gamma}(z_{sub}(1),p_1',k) \vec{k} +
 e  \hat{\delta}(z_{sub}(1),p_1',k)
    i \vec{\sigma} \times \vec p_1^{\,\prime} +    
    e  \hat{\epsilon}(z_{sub}(1),p_1',k) i \vec{\sigma_1} \times \vec{k}\Big) 
+(1\leftrightarrow 2)\, .
\label{kap3_spez2}
\end{eqnarray}

The isospin structure of the scalar functions 
$\hat{\alpha}, ..., \hat{\epsilon}$ 
is parametrized as in~(\ref{1_2}), e.g.\ 
\begin{equation}\label{kap3_iso1}
 \hat{\alpha} =  \,\frac{1}{2} ( \alpha^s + \tau_0 \alpha^v)\, .
\end{equation}
Due to the occurrence of singularities in the loop diagrams, 
 $\hat{\alpha}, ..., \hat{\epsilon}$ 
 become complex functions above $\pi$-threshold. 
Therefore, we do not perform a Taylor expansion of these functions
with respect to the arguments $k$ and $p_1'$. The term
 proportional to $\hat{\alpha}$ can be interpreted as an electromagnetic
 loop correction
 to the charge of the nucleon. The contributions proportional
 to $\hat{\beta}$ and $\hat{\gamma}$ correspond to the well-known convection-,
 the term proportional to $\hat{\epsilon}$ to the spin-current.

Now we will fix the remaining contribution $J_{1}^{N[1]\, \mu}(z,\vec{k})$.
As already mentioned above, $J_{1}^{N[1]\, \mu}(z,\vec{k})$ consists of 
two parts
\begin{equation}\label{3_22}
J_{1}^{N[1]\, \mu}(z,\vec{k}) = \sum_{i=1,2}
 j^{\mu}_{bare}(i,\vec{k}) 
 + J^{\mu}_{counter}(\vec{k})\,, 
\end{equation}
where $j^{\mu}_{bare}$ describes the one-body current of a bare nucleon 
with the physical charge $\hat{e}$ and a vanishing anomalous magnetic moment.
 It is given by  
\begin{eqnarray}
 \bra{{\bar N}(\vec{p}^{\,\prime})}
 \rho_{bare}
\ket{{\bar N}(\vec{p}\,)}  &=&
\delta(\vec{p}^{\,\prime}-\vec{p}-\vec{k}) \,\hat{e}
 \, , \label{kap3_bare_ladung} \\
 \bra{{\bar N}(\vec{p}^{\,\prime})} \vec{\jmath}_{bare}
\ket{{\bar N}(\vec{p}\,)}  &=&
\delta(\vec{p}^{\,\prime}-\vec{p}-\vec{k}) \left\{
 \frac{\hat{e}}{2M_N} (\vec{p}^{\, \prime} +\vec{p}\,) +
 \frac{\hat{e}+\hat\kappa_{bare}}{2M_N} \,i\,\vec{\sigma}\times\vec{k}
 \right\}\,,  \label{kap3_bare_strom}
\end{eqnarray}
 where $\hat{\kappa}_{bare} \equiv 0$. 
As usual, the isospin structure of the charge operator 
$\hat{e}$   is  parametrized as
\begin{equation}\label{3_bar_e}
 \hat{e}=\frac{e}{2} (1 + \tau_0 ) \,\, . 
\end{equation}

Concerning the additional contribution $J^{\mu}_{counter}$, the simplest 
treatment  obviously would be to set
 $J^{\mu}_{counter} \equiv 0$. By this choice, 
the meson cloud contribution (\ref{kap3_spez2}) would change  
the charge of the nucleon and
 would generate an  anomalous magnetic moment which is known to be unrealistic.
 Moreover,  one can show that with this choice  
gauge invariance is not fulfilled even approximately in the leading order
  $1/M_N$ 
(see the discussion below). Therefore, we fix the counter current 
by requiring that the {\it onshell} matrix element of the  effective 
one-body nucleonic current $J_{eff}^{N[1]\, \mu}(z,\vec{k})$ 
  between plane waves is identical  to  the
 realistic  one-body current $\vec{j}_{real}^{N[1] \mu}$ with the same
  formal  expression as  in~(\ref{kap3_bare_ladung}) and 
(\ref{kap3_bare_strom}) except for a nonvanishing anomalous magnetic 
moment contribution $\hat{\kappa}_{phys}$ of the physical nucleon, i.e.\
\begin{equation}\label{4_kp}
\hat{\kappa}_{bare} \rightarrow \hat{\kappa}_{phys} = \frac{e}{2}
\left( \kappa^s_{phys} + \tau_0 \kappa^v_{phys} \right)
\end{equation}
 in (\ref{kap3_bare_strom}).
In detail, this means that the counter current $J^{\mu}_{counter}$ is
 defined by requiring
\begin{eqnarray}\label{fix_counter}
 \bra{{\bar N} (\vec{p}_1^{\,\prime}\,)\, {\bar N} (\vec{p}_2^{\,\prime}\,)} 
J_{eff}^{N[1]\, \mu}(z^{os},\vec{k})
 \ket{ \bar N (\vec{p}_1\,)\,\,\bar N (\vec{p}_2\,)}  
&=&
 \bra{{\bar N} (\vec{p}_1^{\,\prime}\,)\,\, {\bar N} (\vec{p}_2^{\,\prime}\,)} 
\sum_{i=1,2} j_{real}^{N[1]\, \mu}(i,\vec{k})
 \ket{ \bar N (\vec{p}_1\,)\,\,\bar N (\vec{p}_2\,)}  
\end{eqnarray}
with  $z^{os} := e_N(p_1') + e_N(p_2')$.
We would like to emphasize that this condition allows us to reproduce
 the physical onshell one-body current 
no matter what model is used in estimating the loop contributions.
In other words, specific approximations of the adopted model 
like, for example, the neglect of heavier mesons or of two-loop contributions,
do not affect the  onshell properties of the effective 
current. Only the offshell continuation of the effective
 current is model dependent.
 
A straightforward evaluation  of~(\ref{fix_counter}) yields for the counter 
current the expression 
\begin{eqnarray}
 \bra{{\bar N} (\vec{p}_1^{\,\prime}\,)\, {\bar N} (\vec{p}_2^{\,\prime}\,)} 
&& J^{\mu}_{counter}(\vec{k}) 
 \ket{ \bar N (\vec{p}_1\,)\,\,\bar N (\vec{p}_2\,)}   \nonumber\\
 & &\!\!\!\!\!\!\!\!\!\!\!\!\!\!\!\!\!\!\!\!\!\!\!\!\!\!\!\!\!\!\!
=  \bra{{\bar N} (\vec{p}_1^{\,\prime}\,)\,\, {\bar N} (\vec{p}_2^{\,\prime}\,)} 
 - \sum_{i=1,2} j^{\mu}_{bare}(i,\vec{k}) 
 -  {\cal J}_{loop}^{\mu}(z^{os},\vec{k})
 +      \frac{{\widehat Z}^{os}_{[2]}} {{\widehat R}(z^{os})} 
\sum_{i=1,2} j_{real}^{N[1]\, \mu}(i,\vec{k})
\frac{{\widehat Z}^{os}_{[2]}} {{\widehat R}(z^{os}-k)} 
 \ket{ \bar N (\vec{p}_1\,)\,\,\bar N (\vec{p}_2\,)}  \nonumber \\ & &  
\label{3_counter}
\end{eqnarray}
Collecting all pieces, the matrix element of the effective 
nucleon one-body operator between two-body plane-waves
 can then be written in the form
\begin{eqnarray}
 \bra{{\bar N} (\vec{p}_1^{\,\prime}\,)\, {\bar N} (\vec{p}_2^{\,\prime}\,)} 
&&J_{eff}^{N[1]\, \mu}(z,\vec{k})
 \ket{ \bar N (\vec{p}_1\,)\,\,\bar N (\vec{p}_2\,)}   \nonumber\\
 & &\!\!\!\!\!\!\!\!\!\!\!\!\!\!\!\!\!\!\!\!\!\!\!\!\!\!\!\!\!\!
= \bra{{\bar N} (\vec{p}_1^{\,\prime}\,)\, {\bar N} (\vec{p}_2^{\,\prime}\,)} 
 \frac{{\widehat R}(z)}{{\widehat R}(z^{os})}
 \sum_{i=1,2} j_{real}^{N[1]\, \mu}(i,\vec{k})
 \frac{{\widehat R}(z-k)}{{\widehat R}(z^{os}-k)} +
 {\widehat R}(z) {\cal J}_{loop,\, sub}^{\mu}(z,\vec{k})  {\widehat R}(z-k)
 \ket{ \bar N (\vec{p}_1\,)\,\,\bar N (\vec{p}_2\,)} \,\, .   
 \label{3_final}
\end{eqnarray}
 Note that, according to (\ref{5_r1}) through (\ref{4_ss}),
 the plane-wave matrix element of 
  ${\widehat R}(z^{os})$ is identical to one (apart from the  momentum
 conserving Delta-function). The matrix element of the current 
${\cal J}_{loop,\,sub}^{N\mu}(z,\vec{k})$ is obtained from 
(\ref{kap3_spez1}) and  (\ref{kap3_spez2}) by the replacements
\begin{eqnarray}
\hat{x}(z_{sub},p,k) & \rightarrow&
\hat{X}(z_{sub},p,k) :=
\hat{x}(z_{sub},p,k) -\hat{x}(z^{os}_{sub} := e_N(p),p,k)  
\label{3_ersetz}
\end{eqnarray}
for  $x \in \{ \alpha,\beta,\gamma,\delta, \epsilon \}$ and  
 $X  \in \{A,B,C,D,E \} $, 
where the isospin structure of $\hat{A},...,\hat{E}$  is parametrized 
in analogy to (\ref{kap3_iso1}) introducing isoscalar and isovector 
parts $A^s,...,E^s$ and $A^v,...,E^v$, respectively.

We would like to remind the reader that offshell effects arise from 
two sources in our model. The first one comes from the purely hadronic 
renormalization factors $\widehat{R}$ in the initial and final states and 
the second one from the loop contribution to the electromagnetic current.
In the usual treatment without offshell effects like in \cite{ScA00}, one
neglects in (\ref{3_final}) the term ${\cal J}_{loop,\, sub}^{\mu}$
 and substitutes ${\widehat R}$  by one.

As last topic in this section we will study the question of gauge invariance. 
By a straightforward calculation, one obtains for the loop contributions 
($\rho \equiv J^0$)
\begin{eqnarray}
 \vec{k} \cdot 
\sum_{j=2}^9 \vec{J}_{j}^{N[1]}(z,\vec{k})
  &=&  \sum_{i=1,2} V^{\pi\,dis}_{eff}(i,z) {\rho}_{real}(i,\vec{k}) -
 \sum_{i=1,2}
  {\rho}_{real}(i, \vec{k}) V^{\pi\,dis}_{eff}(i,z-k) \nonumber\\  & & + 
  k  \sum_{j=2}^9 \rho_{j}^{\,N[1]}(z,\vec{k})
  + {\cal B}(z_f, \vec{k}\,)
  + {\hat {\cal O}}\left(\frac{1}{M_{N}^3}\right)\,,
 \label{kap5_kontin_pi_self_1}
\end{eqnarray}
 with the retarded disconnected  $\pi$-loop contribution
\begin{equation}\label{kap5_kontin_pi_pot_1}
 V^{\pi\,dis}_{eff}(i,z) = 
  v^{0}_{{\bar N} X_{\pi}}(i) \, G_0(z) 
 v^{0}_{X_{\pi}{\bar N}}(i) \, 
\end{equation}
 and  the auxiliary quantity
 \begin{eqnarray}
 {\cal B}(z, \vec{k}\,) &=&
  \sum_{i=1,2}\, 
 \left\{ \left(
  v^{0\, nonrel}_{{\bar N} X_{\pi}}(i)
  G_0(z) v^0_{X_{\pi} X_{\pi \pi}}(i)  G_0(z)
  \, {\rho}^{(0)}_{X_{\pi \pi} {\bar N} }(\vec{k}) \right)
  \, \left(z-k - H_{0\,{\bar N}{\bar N}} \right) \right.\nonumber\\
  & & \left. \quad -
 \left(z - H_{0\, {\bar N}{\bar N}} \right) \,
  \left(
   \, {\rho}^{(0)}_{{\bar N} X_{\pi \pi}} (\vec{k})\,
 G_0(z-k)
  v^{0\, nonrel}_{X_{\pi \pi}X_{\pi}}(i) G_0(z-k) v^{0}_{X_{\pi}
  {\bar N}}(i) \right) \right\}\, . 
 \label{kap3_b}
\end{eqnarray}
In analogy to the corresponding discussion for the retarded MEC 
in~\cite{ScA00}, we would like to mention that the matrix element of 
${\cal B}(z, \vec{k}\,)$ vanishes between plane waves only, which fact 
indicates that additional loop contributions of at least fourth order in 
the pion-nucleon coupling 
are needed in order to fulfill gauge invariance. 
 As has been indicated in  (\ref{kap5_kontin_pi_self_1}), there exist
 additional sources for the breaking of gauge invariance in the
 order  $1/M_N^3$.
 They have their origin in the fact that in the loop corrections 
 the nonrelativistic expressions for the pion-nucleon vertex 
  and for the momentum-energy relation 
of the intermediate nucleon are used whereas in   $V^{\pi\,dis}_{eff}(z)$,
 which enters into our hadronic interaction (Elster potential, see
 \cite{ScA99} for further details), the relativistic expressions have been
 kept.

In our explicit evaluations, we have used the Siegert decomposition of the
electric multipoles with the help of the relation 
\begin{equation}
  \bra{\alpha_{W}} 
 \vec{k} \cdot\left[ 
\vec{J}_{eff}^{\,N[1]}(z,\vec{k})
+ \vec{\cal{J}}_{eff}^{\,N[2]}(z,\vec{k}) \right] 
 \ket{\alpha_{W-k}}  =  
\bra{\alpha_{W}}
  k  \left[  {\cal J}_{eff}^{N[1]\,0}(z,\vec{k}) +  
   {\cal J}_{eff}^{N[2]\,0}(z,\vec{k})  \right] 
\ket{\alpha_{W-k}} \,\, ,  \label{kap3_siegert} 
\end{equation}
where $z = W + i \epsilon$.
 Here, ${\cal J}_{eff}^{N[1]\,\mu}(z,\vec{k})$ is given by
 (\ref{3_final}) and ${\cal J}_{eff}^{N[2]\,0}(z,\vec{k})$ and 
$\vec{\cal{J}}_{eff}^{\,N[2]}(z,\vec{k})$ denote the retarded two-body
charge and currents given in equation (40) of  \cite{ScA00}, 
 multiplied in 
 addition by  ${\widehat R}(z)$ from the left and
 by ${\widehat R}(z-k)$ from the right according to  (\ref{change}). 
 If the $\Delta$-isobar 
as well as the interaction $V^0_{XX}$ are ignored and if only pions are 
considered as explicit mesonic degrees of freedom, (\ref{kap3_siegert}) 
can be derived rigorously in the leading order $1/M_N$ exploiting
(\ref{kap5_kontin_pi_self_1}) and the similar expression for the retarded 
MEC of~\cite{ScA00}.
 However, we use the Siegert decomposition
 (\ref{kap3_siegert}) also in our 
full model. Thereby, as it is well-known, at least a part of 
the neglected currents like heavy-meson exchange contributions can be taken 
into account implicitly. We would like to emphasize that 
(\ref{kap5_kontin_pi_self_1}) is violated even in the leading order 
$1/M_N$ if the counter current $J^{\mu}_{counter}$ is set equal to zero.
 
Finally, we would like to  remark that of the currents of leading 
 relativistic order,  the important spin-orbit current \cite{CaM82,WiL88}
  is of course included in our numerical evaluation.

\section{Results for the loop contributions}\label{kap4}
\begin{figure}[ppp]
\centerline{\psfig{figure=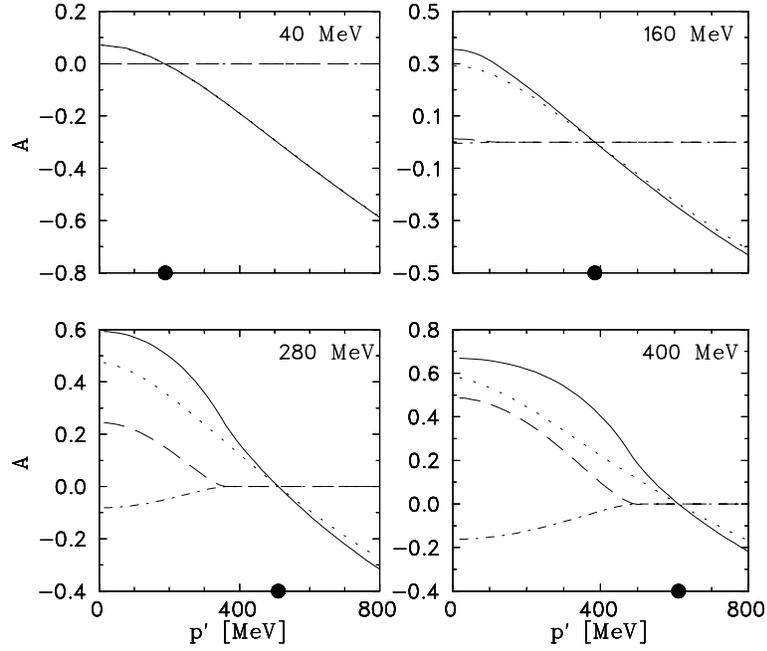,width=10cm,angle=0}}
\vspace{0.5cm}
\caption{
The isospin components of the
 function $\hat{A}(p',k)$ (\ref{3_ersetz}) of the effective loop current
 ${\cal J}_{loop,\,sub}^{N\mu}(z,\vec{k})$
 for various laboratory
 energies $k_{lab}$ of the incoming photon. Notation of the curves:
 solid: real  part $A^s$, dashed: imaginary  part  $A^s$,
 dotted: real part  $A^v$,  dash-dotted:  imaginary part $A^v$.
 The location of the 
 onshell momentum $p_0$ of the outgoing nucleon, defined in (\ref{1_5}), 
 is indicated by a filled circle.}
\label{fig_a}
\end{figure}

\begin{figure}[ppp]
\centerline{\psfig{figure=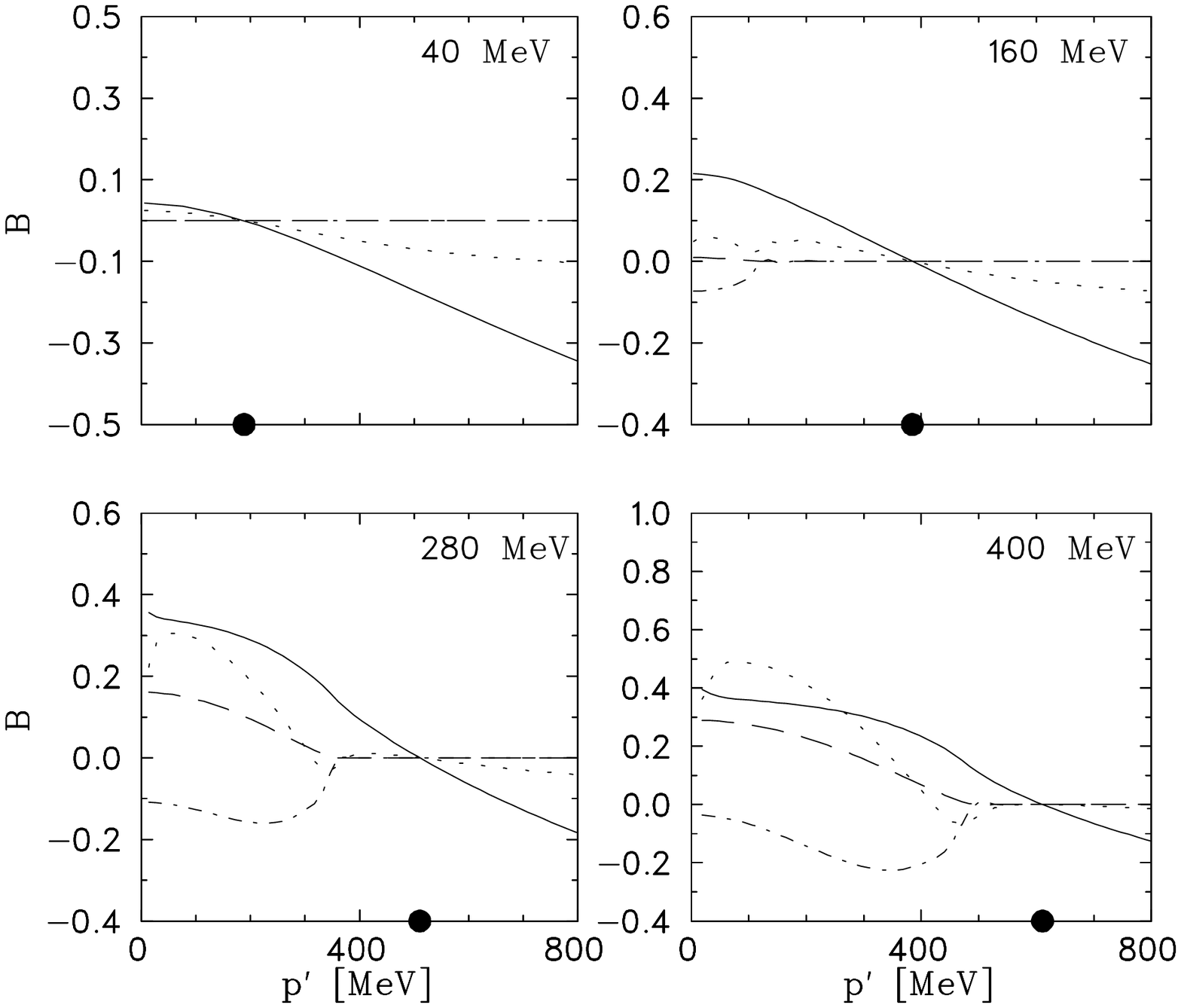,width=10cm,angle=0}}
\vspace{0.5cm}
\caption{
The  isospin components of the function $\hat{B}(p',k)$ (\ref{3_ersetz}) 
of the effective loop current  for various laboratory
 energies $k_{lab}$ of the incoming photon. Notation of the curves
 in analogy to Fig. \ref{fig_a}.}
\label{fig_b1}
\end{figure}

\begin{figure}[ppp]
\centerline{\psfig{figure=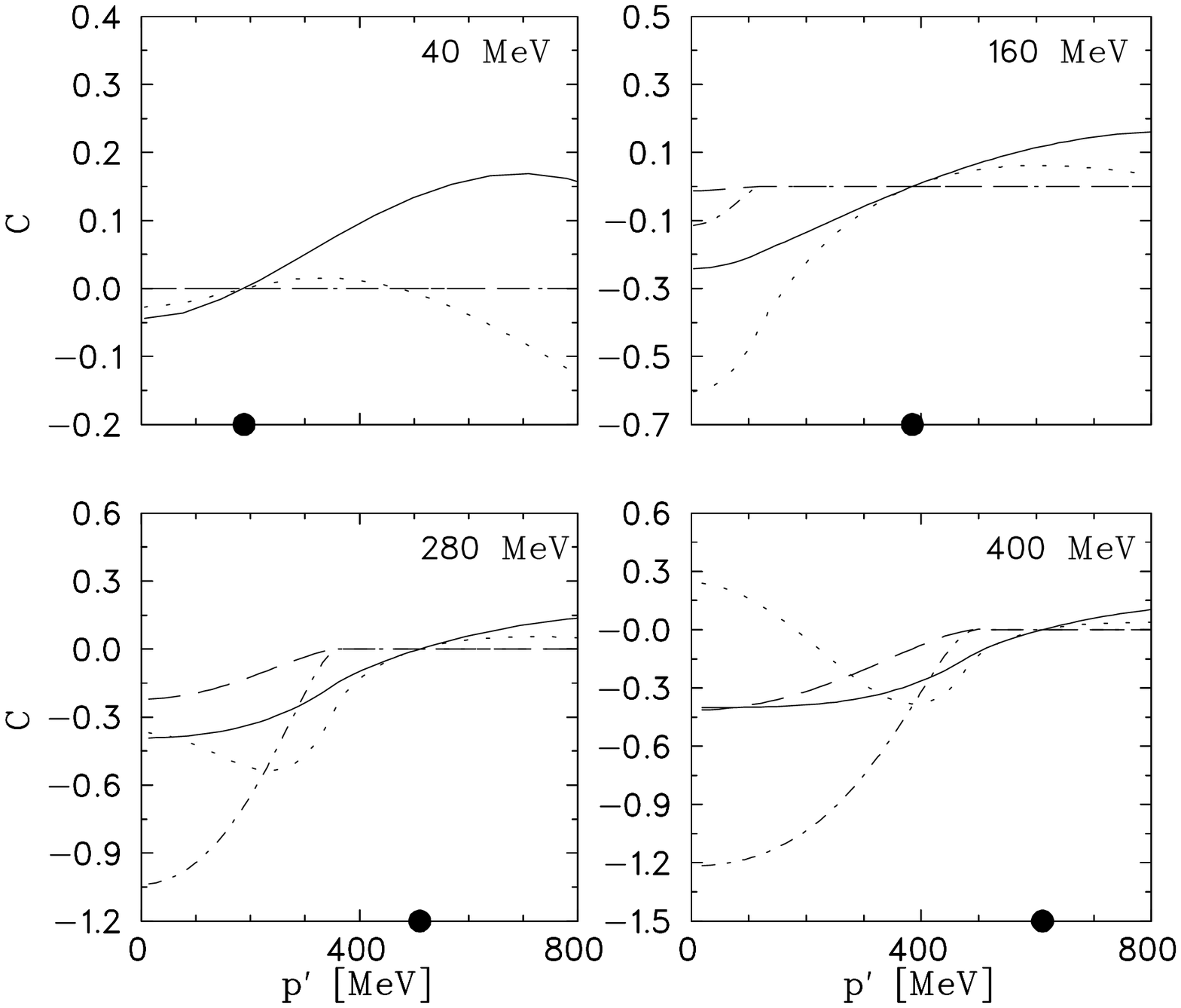,width=10cm,angle=0}}
\vspace{0.5cm}
\caption{
The   isospin components of the 
 function $\hat{C}(p',k)$  (\ref{3_ersetz}) of the effective loop current
 for various laboratory
 energies $k_{lab}$ of the incoming photon. Notation of the curves
 in analogy to Fig.\ \ref{fig_a}.}
\label{fig_b2}
\end{figure}

\begin{figure}[ppp]
\centerline{\psfig{figure=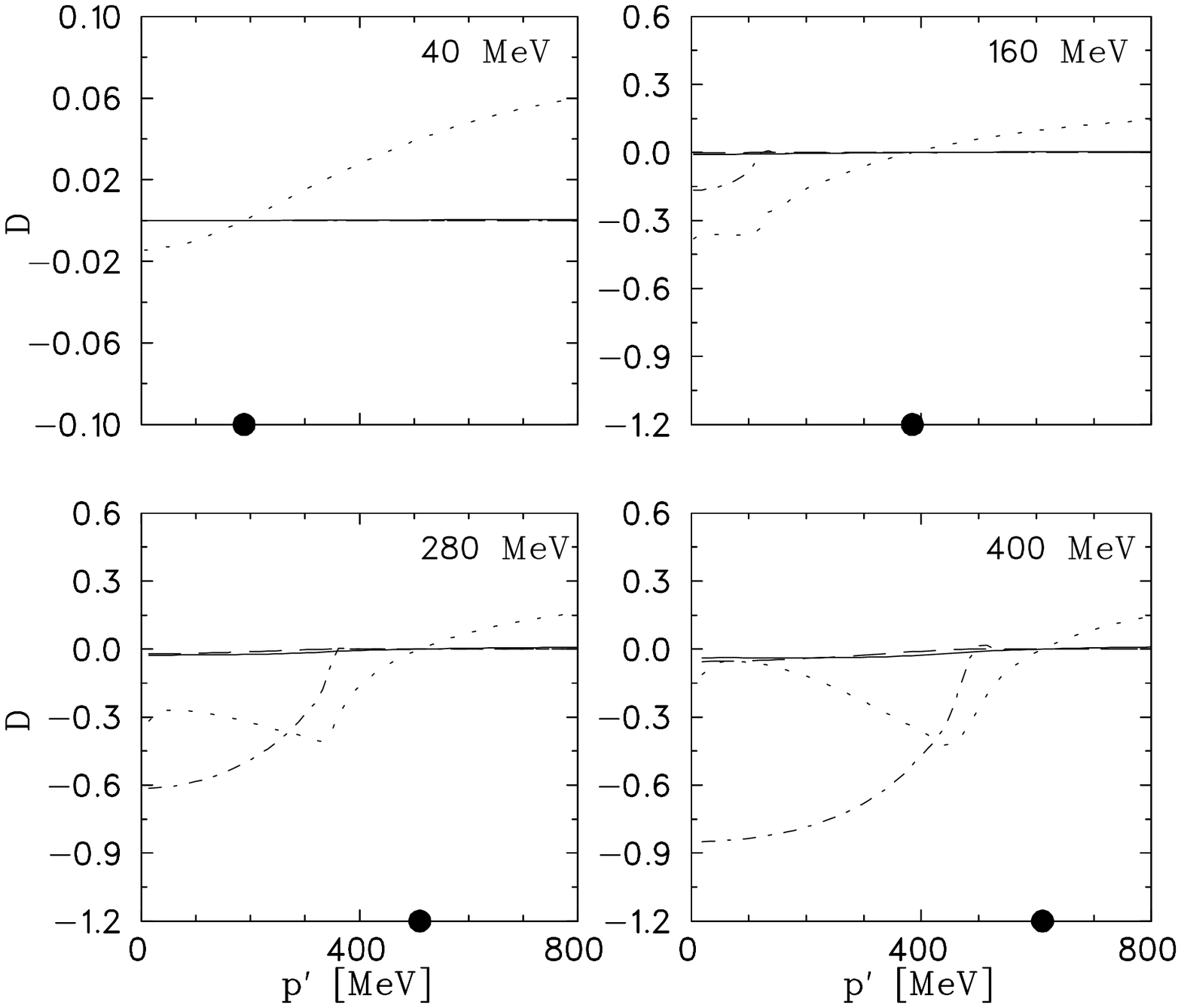,width=10cm,angle=0}}
\vspace{0.5cm}
\caption{
The isospin components of the 
  function $\hat{D}(p',k)$ (\ref{3_ersetz}) of the effective loop current
 for various laboratory
 energies $k_{lab}$ of the incoming photon. Notation of the curves
 in analogy to Fig.\ \ref{fig_a}.}
\label{fig_b3}
\end{figure}

\begin{figure}[ppp]
\centerline{\psfig{figure=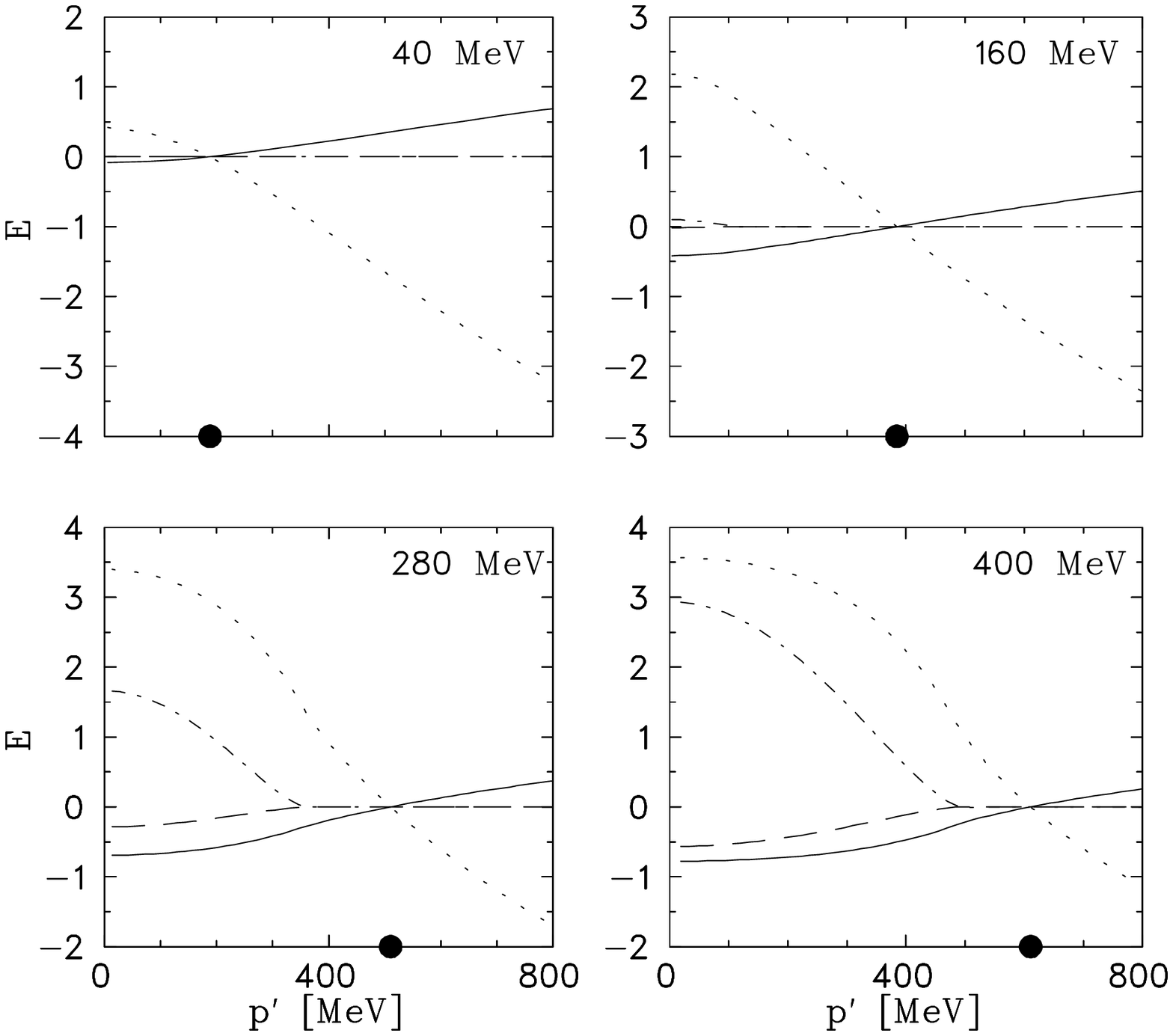,width=10cm,angle=0}}
\vspace{0.5cm}
\caption{
The isospin components of the 
  function $\hat{E}(p',k)$ (\ref{3_ersetz}) of the effective loop current
 for various laboratory
 energies $k_{lab}$ of the incoming photon. Notation of the curves
 in analogy to Fig.\ \ref{fig_a}.}
\label{fig_b4}
\end{figure}

For the explicit calculation, we have chosen a reference frame where
the two baryons have total momentum zero after the absorption of the photon.
  To begin the discussion we present in 
Figs.~\ref{fig_a} through \ref{fig_b4} our results for the quantities 
$\hat{A}(z_{sub},p',k),...,\hat{E}(z_{sub},p',k)$ of 
 (\ref{3_ersetz}), 
which enter into the effective current in (\ref{3_final})
 via the contribution ${\cal J}_{loop,\,sub}^{N\mu}(z,\vec{k})$.
  In this context, 
one should remember that $p'$ is the absolute value of the struck
 nucleon momentum
after the absorption of the photon. Due to the chosen reference frame, 
$-\vec{p}^{\,\prime}$ is the 
momentum of the spectator nucleon, so that $z_{sub}$ 
 is not an  independent variable but determined by $p'$ and $k$: 
\begin{equation}\label{5_zsub}
z_{sub}\equiv z_{sub}(k,p')= W  - \sqrt{M_N^2 + {p'}^2} + i \epsilon\,\, ,  
\end{equation}
 with the invariant energy
\begin{equation}\label{1_5b}
W=\sqrt{M_d^2 + 2 M_d k_{lab}}\,,
\end{equation}
where $M_d$ and  $k_{lab}$ denote the deuteron mass and 
 the lab photon energy, respectively. The latter is related to $k$,
 the c. m.\  photon energy, via  
\begin{equation}\label{1_5c}
k = k_{lab} \frac{M_d}{W} \,\, .
\end{equation}
 Therefore, the functions 
$\hat{A}(z_{sub},p,k),...,\hat{E}(z_{sub},p,k)$
 can be treated effectively as functions solely of $k$ and $p'$.  
 For later purposes, we introduce in addition 
 the so-called pole momentum $p_0$ via
\begin{equation}\label{1_5} 
W = 2 \sqrt{M_N^2+ p_0^{\,2}}\,. 
\end{equation}
It is identical to the asymptotic  relative momentum of the outgoing nucleons 
in photodisintegration. Due to (\ref{3_ersetz}), the functions
 $\hat{A},...,\hat{E}$ are equal to zero at $p_0$. Moreover, they
 vanish for $k$ and $p' \rightarrow 0$. 
Therefore, as we will see in the next section, the loop contributions do not
play any role in the observables at low energies small compared to
 the  $\pi$-threshold.
For increasing energy, the analytical behavior of $\hat{A},...,\hat{E}$
as a function of $k$ shows in general a more and more pronounced slope. 
Beyond $\pi$-threshold, there exists for momenta $p' < p_{max}(k)$ for 
certain values of $p_{max}$ smaller than $p_0$ an imaginary part due 
to occurring 
singularities in the pion-loops. In the $\Delta$-region, its absolute size 
is in general of the same order of magnitude than the corresponding real parts.
Moreover, for small $p'$ the 
absolute size of the functions depicted in Figs.~\ref{fig_a} through 
\ref{fig_b4} is comparable to the corresponding values (\ref{1_3}) of the 
charge and the anomalous magnetic moment of the nucleon.

\begin{figure}[ppp]
\centerline{\psfig{figure=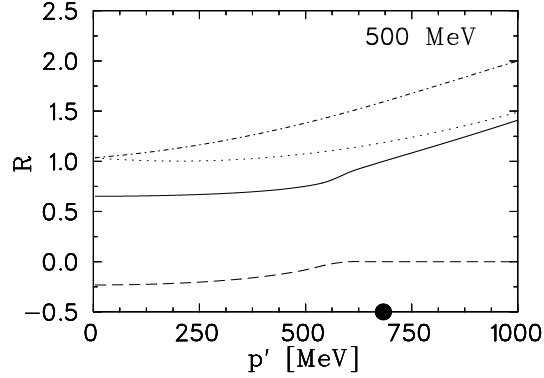,width=7cm,angle=90}}
\vspace{0.5cm}
\caption{The matrix element (\ref{4_rr2})  of the dressing factor  
 for $k_{lab}=500$ MeV. Notation of the 
 curves: solid: real part of $R_f(k,p')$, dashed: imaginary part of
 $R_f(k,p')$, dotted: $R_i(z-k,p',k,x=1)$, dash-dotted: $R_i(k,p',x=-1)$.
 $R_i$ has no imaginary part.
 The location of the onshell momentum $p_0$ 
of the outgoing nucleon 
 is indicated by a filled circle.}
\label{fig_rr}
\end{figure}

Besides the electromagnetic loop contributions 
${\cal J}_{loop,\,sub}(z,\vec{k})$ entering (\ref{3_final}), 
an additional new feature of our effective one-body current, compared to 
the conventional onshell expression, is the occurrence of the 
dressing operator ${\widehat R}$, appearing before ($i$) and after ($f$) 
 photon 
absorption and described according to (\ref{5_r1}) by 
$R_i(z,\vec p_1,\vec p_2)$ and 
 $R_f(z,\vec p^{\,\prime}_1,\vec p^{\,\prime}_2)$, respectively.
Whereas $R_f= R_f(k,p')$ depends only on $k$ and $p'$, because 
$p'_1 = p'_2 \equiv p'$ and $z= \sqrt{M_d^2 + k^2} + k + i \epsilon$, one 
finds that $R_i= R_i(k,p',x)$
depends also  via $x = \hat{p'}\cdot\hat k$ on 
the angle between $\hat{p'}$ and $\hat{k}$, because the deuteron is not 
in its rest frame. 
 In general, as becomes evident 
 in Fig.~\ref{fig_rr}, one notes that  $R_f(k,p')$ is smaller or larger than
1 for momenta $p'$ smaller or larger than the onshell value 
$p_0$, respectively, whereas $R_i(k,p',x)$ in the initial state is always 
larger than 1, leading therefore to a considerable 
amplification of the loop contribution
${\cal J}_{loop,\,sub}(z,\vec{k})$ in  (\ref{3_final}). 
In this context we would like to point out another interesting feature of 
the dressing factors, namely if the outgoing nucleons are onshell,  
 i.e.\ $R_f(k,p_0) =1$, the value of the corresponding dressing factor 
$R_i(k,p_0,x)$  is not equal to one, indicating that one of the initial 
nucleons has to be offshell. This behavior can well be understood
by remembering that the absorption of a {\it real} photon by a free nucleon 
is not possible without internal excitation. 
 
\begin{figure}[ppp]
\centerline{\psfig{figure=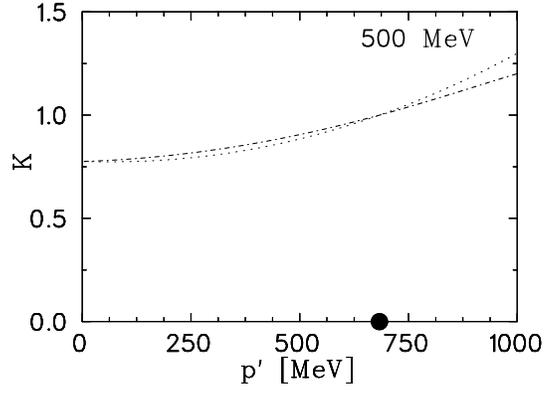,width=7cm,angle=90}}
\vspace{0.5cm}
\caption{The function $K$ (\ref{5_rrr}) for $k_{lab}=500$ MeV. 
Notation of the 
 curves: dotted: $K(z-k,p',k,x=1)$, dash-dotted: $K(z-k,p',k,x=-1)$.
 $K$ has no imaginary part.
 The location of the onshell momentum $p_0$ of the outgoing nucleon 
 is indicated by a filled circle.}
\label{fig_rr0}
\end{figure}

Finally, we would like to discuss the matrixelement of
 the quantity ${\widehat R}(z-k)/{\widehat R}(z^{os}-k)$, i.e.\
\begin{equation}\label{5_rrr}
 \bra{{\bar N} (-\vec{p}^{\,\prime}\,)\,\, {\bar N} (\vec{p}^{\,\prime}
 - \vec{k}\,)} 
 \frac{{\widehat R}(z-k)}{{\widehat R}(z^{os}-k)}
 \ket{ \bar N (-\vec{p}\,)\,\,\bar N (\vec{p} - \vec{k}\,)}  \nonumber\\
 = \delta(\vec{p}^{\,\prime}-\vec{p}) \,\,
  K(z-k,p', k,\hat{p'}\cdot \hat{k})\,\, ,
\end{equation}
which appears in (\ref{3_final}). According to Fig.\   \ref{fig_rr0}, $K$ 
is smaller than 1 in the  most relevant
 momentum range $p' < p_0$, leading therefore to a
 weakening of the contribution of $j^{\,nr,\, \mu}_{real}$ in (\ref{3_final}).
  In contrast to
 $R_i$, the function $K$ is almost independent of 
 $x = \hat{p'}\cdot \hat{k}$.

\section{Offshell effects in deuteron photodisintegration}\label{kap5}

To begin with, we will describe very briefly the ingredients
 of the adopted hadronic interaction 
model for which we use the model 
``CC(ret2)'' of \cite{ScA00} as our reference point for offshell effects. 
It incorporates full retardation in potentials and two-body currents,
$\pi$- as well as $\rho$-exchange in the $N\Delta$-interaction and the 
coupling to the $\pi d$-channel. For the effective nucleon one-body current, 
the usual onshell expression including spin-, convection- and the 
spin-orbit currents. Any offshell effects from the electromagnetic pion-loops 
or from the dressing operator are neglected. 

\begin{figure}[ppp]
\centerline{\psfig{figure=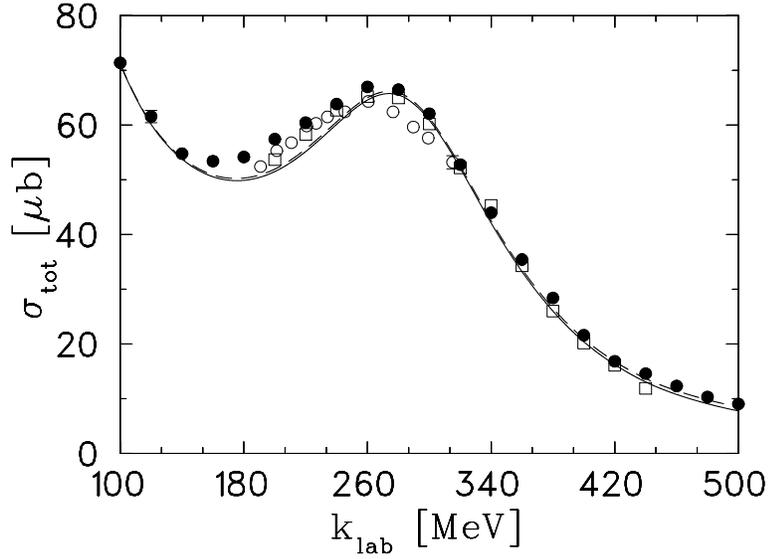,width=10cm,angle=90}}
\vspace{0.5cm}
\caption{Total cross section $\sigma_{tot}(\gamma d \rightarrow NN)$
 of deuteron photodisintegration. Notation
 of the curves: dashed: $\sigma_{\mathrm{tot}}^{\mathrm{onsh}}$,
 solid: $\sigma_{\mathrm{tot}}^{\mathrm{offsh}}$. Experimental data 
from {\protect \cite{CrA96}} ($\bullet$), 
 {\protect \cite{ArG84}} ($\Box$) and
{\protect \cite{BlB95}} ($\circ$).         }
\label{fig_wqtot_pd}
\end{figure}

\begin{figure}[htp]
\centerline{\psfig{figure=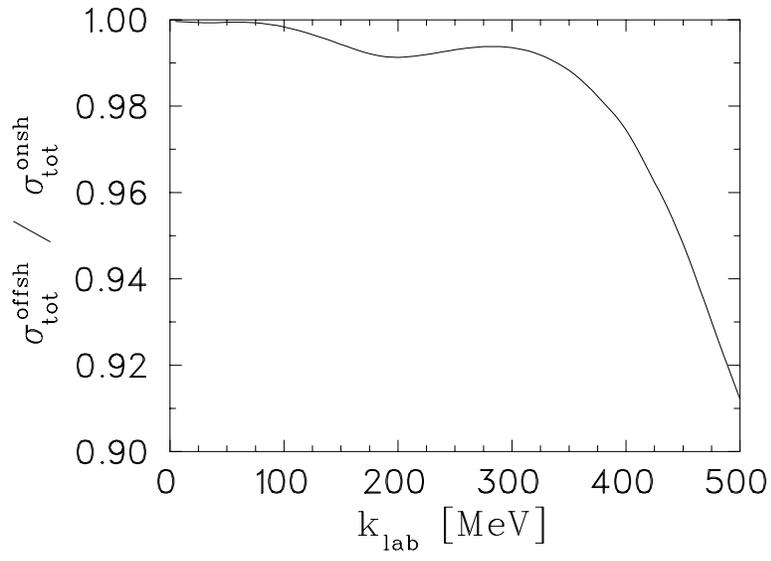,width=10cm,angle=90}}
\vspace{0.5cm}
\caption{Ratio      
  $\sigma_{\mathrm{tot}}^{\mathrm{offsh}}/\sigma_{\mathrm{tot}}^{\mathrm{onsh}}$ 
 of the total cross section 
 of deuteron photodisintegration.    }
\label{fig_wqtot_verh}
\end{figure}

\begin{figure}[ppp]
\centerline{\psfig{figure=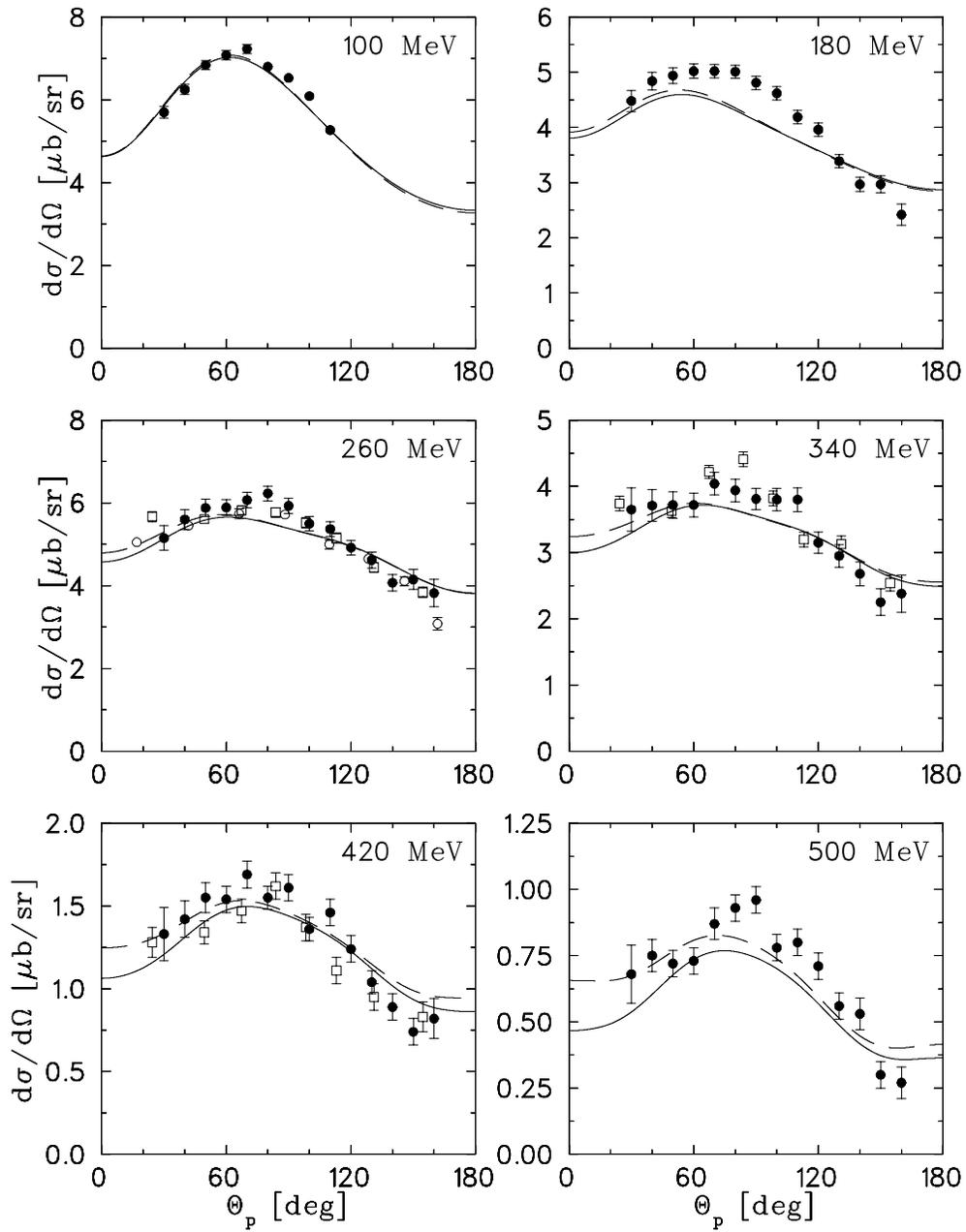,width=13cm,angle=0}}
\vspace{0.5cm}
\caption{Differential cross sections of deuteron photodisintegration for 
various energies. Notation of the curves and experimental data as in 
Fig.~\ref{fig_wqtot_pd}.}
\label{fig_wqdiff_pd}
\end{figure}

\begin{figure}[ppp]
\centerline{\psfig{figure=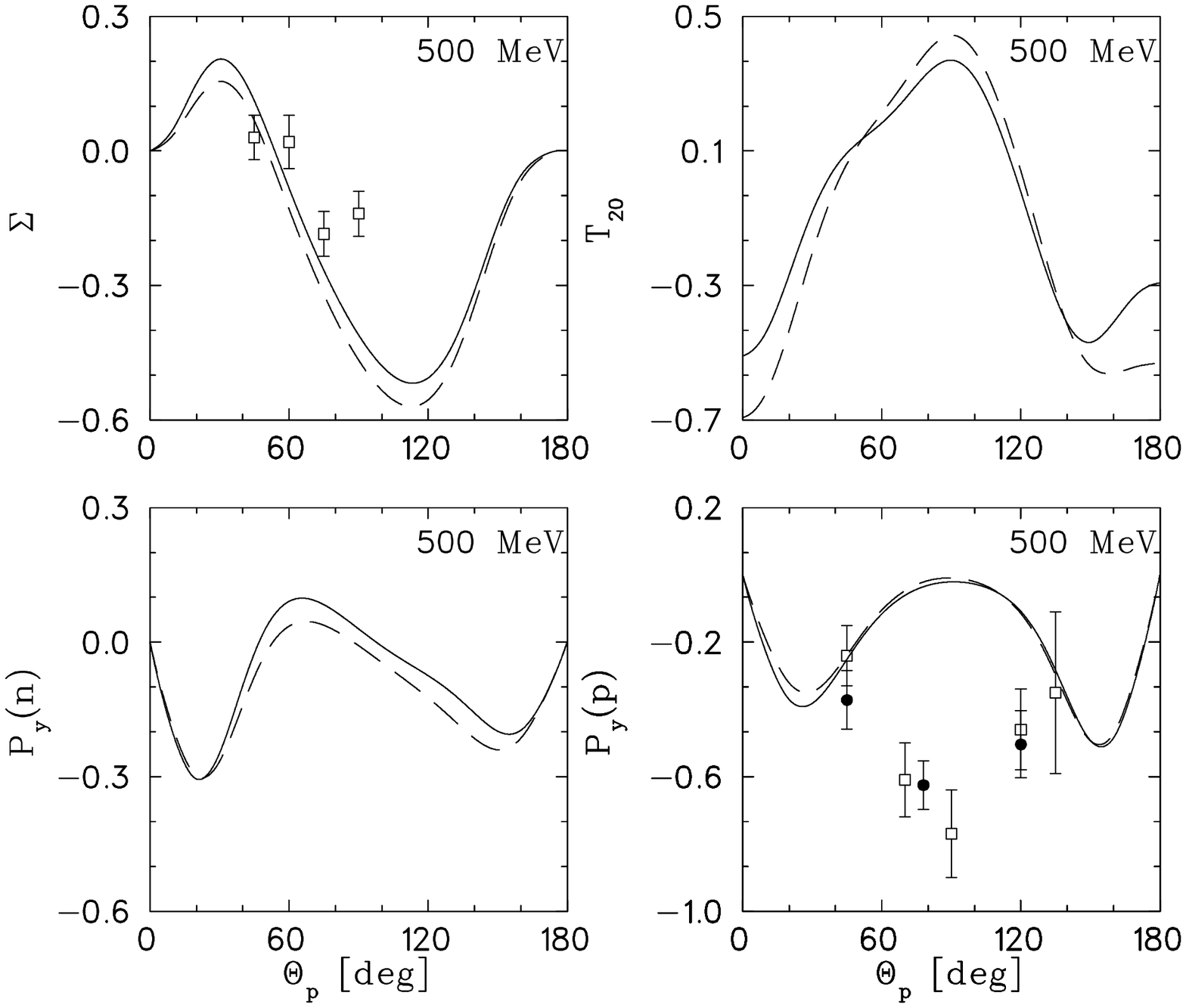,width=13cm,angle=0}}
\vspace{0.5cm}
\caption{Linear photon asymmetry $\Sigma$, tensor target asymmetry $T_{20}$, 
neutron polarization $P_y(n)$
 and proton polarization $P_y(p)$ in deuteron photodisintegration
 for $k_{lab}=500$ MeV. Notation of the curves  as in 
Fig.~\ref{fig_wqtot_pd}.
 Experimental data for $\Sigma$ from {\protect  \cite{AdB91}} ($\Box$),
 for $P_y(p)$ from {\protect \cite{IkA79}} ($\Box$) and 
 {\protect \cite{BrD80}} ($\bullet$).}
\label{fig_sigma_neupol_pd}
\end{figure}

We now turn to the discussion of our numerical results. We indicate by a 
superscript ``onsh'' or ``offsh'' in the captions of 
Figs.~\ref{fig_wqtot_pd} through \ref{fig_sigma_neupol_pd} whether the 
considered observable is calculated using the conventional onshell current
or in the offshell approach (\ref{3_final}) for the effective one-body 
current, respectively. First, we will consider the total cross section for 
photodisintegration presented in Fig.~\ref{fig_wqtot_pd}. In order to 
show more clearly the relative offshell effect, we show in 
Fig.~\ref{fig_wqtot_verh} the ratio 
$\sigma_{\mathrm{tot}}^{\mathrm{offsh}}/\sigma_{\mathrm{tot}}^{\mathrm{onsh}}$. 
One readily notes that at low energies up to about $k_{lab}=100$~MeV 
offshell effects are almost negligible. Even in the $\Delta$ region,
 up to about 
$k_{lab}=300$~MeV, they do not exceed one percent, because the one-body 
 nucleon current is of minor importance and offshell effects in the
 ${\bar N}\Delta$-excitation are already included in the reference
 calculation \cite{ScA00}. Thus only beyond the 
$\Delta$-peak offshell effects become more and more pronounced, leading to 
a sizeable reduction of about 10 percent at $k_{lab}=500$~MeV. In  
the differential cross sections in Fig.~\ref{fig_wqdiff_pd} the offshell 
effects are most prominent at forward angles decreasing it up to about 
30 \% for the highest energy shown, whereas around $90^{\circ}$ and at backward
 angles the effects 
are much smaller. Our detailed analaysis shows that this result is
 the combined effect of various interferences. It turns out that
 the strong decrease at $0^{\circ}$ for higher energies
 is mainly due to a constructive inferference of the contributions
 proportional to $\hat{A}$ (offshell correction to the charge)
 and $\hat{D}$ (proportional to $\vec{\sigma} \times \vec{p}^{\,\prime}$)
 whereas the other  offshell corrections are of minor importance. On the
 other hand, at $90^{\circ}$ the multiplication of  $j^{N[1]\, \mu}_{real}$
 by the factor 
${\widehat R}(z-k) / {\widehat R}(z^{os}-k)$ in (\ref{3_final})
 leads to a considerable decrease of the cross section, which is, however, 
 partially canceled by the offshell correction $\hat{A}$  to the charge.
  In contrast to the differential cross section, the most interesting 
 single particle polarization 
observables like the photon asymmetry $\Sigma$,
 the tensor target asymmetry $T_{20}$ and the outgoing nucleon 
polarizations $P_y(n)$  and $P_y(p)$ 
  are only moderately affected by offshell effects 
 as is demonstrated in Fig.~\ref{fig_sigma_neupol_pd}. Therefore,
 the incorporation of offshell effects in our framework
  does not resolve the discrepancy between theory and experiment in 
 the  proton polarization around 90$^{\circ}$.

 From these results, 
  we may conclude that for an only  qualitative understanding
of deuteron photodisintegration, one may neglect offshell 
effects safely. However, for a precise theoretical interpretation of 
experimental data with an accuracy on the level of a few percent it is 
certainly necessary to take into account offshell effects as discussed 
in this work. 

\section{Summary and Outlook}\label{kap6}
 In the present paper, we have investigated  the
importance of  offshell effects in deuteron photodisintegration
by evaluating one-pion loop contributions within a hadronic interaction 
model using mesons, nucleons, and $\Delta$-isobars as effective degrees of 
freedom as developed in \cite{ScA99,ScA00}. It is based on a conventional 
nonrelativistic time-ordered perturbation theory and allows a realistic 
description of hadronic and electromagnetic reactions on the two-nucleon 
system for energies up to about 500 MeV excitation energy. 

The offshell effects discussed here consist of two parts, which both are 
related to the same basic meson-nucleon interaction mechanism: 

(i) A purely hadronic offshell contribution arising from the presence of 
a ``nucelon-meson cloud'' as described by the ``dressing factor'' ${\hat R}$. It 
has its origin in the fact that the active nucleon, on which the photon is 
absorbed or emitted, is subject to initial and final state interactions in
the presence of other baryons. 

(ii) An electromagnetic offshell contribution arising from the absorption 
or emission of a photon by this meson cloud as described by the loop 
contributions to the e.m.\ current. By the inclusion of a suitably chosen 
counter current it is ensured that this loop contribution vanishes onshell
 so that the correct onshell properties of the physical nucleon are 
automatically preserved.

We would like to emphasize once more again, that such offshell effects 
depend on the hadronic interaction model and, therefore, as such are not 
separately observable. However, as our simple one-loop model for the offshell 
effects has shown, they will influence theoretical predictions. Despite the 
simplicity of the present model, we expect that it is sufficiently realistic 
enough in order to allow a semiquantitative answer to the interesting question 
as to the size of offshell effects in electromagnetic reactions on bound 
systems.

Our results clearly show that for a theoretical understanding of 
 this reaction  within a, say, 20-30 percent accuracy,
 offshell effects can safely be neglected. 
They do not change qualitatively any of 
the observables, which we have studied,   within that margin. On the other 
hand, these effects are not that small compared to the precision of present 
state-of-the-art experiments with an accuracy of better than 5 percent. 
In view of the fact, that
we found within our framework offshell effects in observables of a size 
which lie in this range, we take this as a clear indication that for a 
theoretical interpretation of high precision data of the order of a few 
percent accuracy, offshell effects have to be taken into account consistently.
But it is also clear, that in this case one needs a more realistic model for 
the internal structure of nucleons including a consistent treatment of 
offshell effects. It reamins for future work to apply the present model
 to other e.m.\ reactions on the deuteron like electrodisintegration,
 pionproduction and Compton scattering.

\section*{Appendix A: Some technical details}
\label{appA}
\begin{figure}[ppp]
\centerline{\psfig{figure=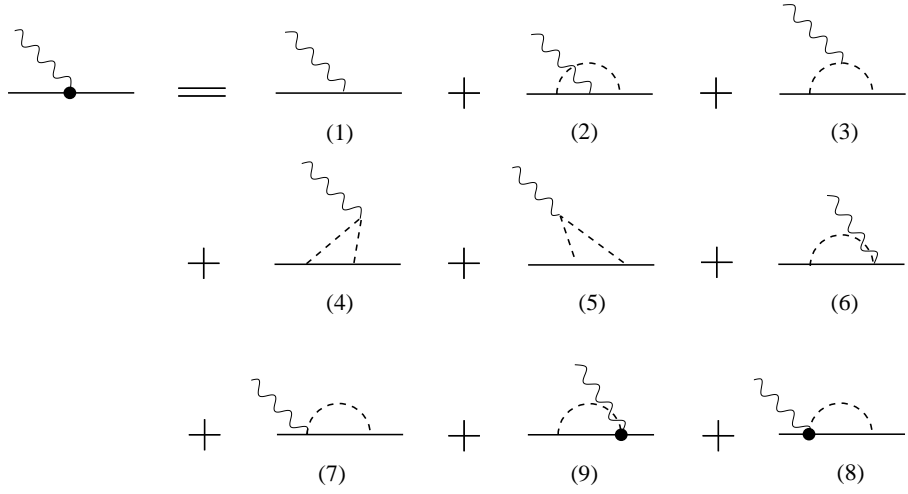,width=12cm,angle=0}}
\vspace{0.5cm}
\caption{
Diagrammatic representation of the separate contributions to 
the effective nucleonic one-body current $J_{eff}^{N[1]\, \mu}(z,\vec{k})$
  (equations (\ref{kap3_i1}) through (\ref{kap3_i9})),
 represented in contrast to the 
bare nucleon current by a filled circle.} 
\label{figem5}
\end{figure}

In this appendix, we illustrate the explicit evaluation of the
 functions $\hat{\alpha},...,\hat{\epsilon}$ in (\ref{kap3_spez1}) and
 (\ref{kap3_spez2}). As an example, we study the contribution
   $J_{6}^{N[1]\, \mu}(z,\vec{k})$ (see (\ref{kap3_i6}) and
 Fig.  \ref{figem5}). First, we split  $J_{6}^{N[1]\, \mu}(z,\vec{k})$
 into  one-body contributions, i.e.\
\begin{equation}\label{app_1}
J_{6}^{N[1]\, \mu}(z,\vec{k})
= \sum_{i=1,2} 
 j_{6}^{N[1]\, \mu}(z,\vec{k},i)
\end{equation}
where,  the argument $i$ denotes the 
 ``active'' nucleon absorbing the photon. The relevant hadronic and
 electromagnetic interactions are given by the renormalized
 nonrelativistic $\pi N$-vertex
\begin{equation}\label{anh1_2}
\bra{{\bar N}(\vec{p}^{\,\prime})} v^{nonrel}_{{\bar N}Q_{\pi}}
 \ket{ \pi(\vec{q},\mu) {\bar N}(\vec{p}\,)}  =
 \delta(\vec{p}^{\,\prime} - \vec{p} - \vec{q}) \,
\frac{g_{\pi}}{2 M_N}
 \tau_{\mu} i   \vec{\sigma} \cdot \vec{q} 
F_{\pi}(\vec{q}^{\,2}) 
\end{equation}
and the Kroll-Rudermann current contribution
\begin{eqnarray}
\bra{\pi(\vec{q},\mu) {\bar N}(\vec{p}^{\,\prime})}
\rho^{(1)}_{Q_{\pi}{\bar N}}(\vec{k}) 
 \ket{ {\bar N}(\vec{p}\,)} &=& 0 \,, \label{anh1_3}\\
\bra{\pi(\vec{q},\mu) {\bar N}(\vec{p}^{\,\prime})}
\vec{\jmath}^{\,\, (1)}_{Q_{\pi}{\bar N}}(\vec{k})
 \ket{ {\bar N}(\vec{p}\,)} &=&
 \delta(\vec{q} - \vec{p} +\vec{k}+ \vec{p}^{\, \prime})\,
 \frac{g_{\pi}}{2M_N}\, (-)^{\mu}\, [\hat{e},   \tau_{-\mu} ]\, 
 i \vec{\sigma} F_{\pi}(\vec{q}^{\,2}) \, \, ,  \label{anh1_4}
\end{eqnarray}
 where $\mu$ denotes the charge of the pion and $\tau_{\mu}$ are the
 usual Pauli-matrices in the spherical basis. 
The hadronic form factor
\begin{equation}\label{anh1_5}
F_{\pi}(\vec{q}^{\,2}) 
= \left( \frac{\Lambda_{\pi}^2 - m_{\pi}^2}{
\Lambda_{\pi}^2 + \vec{q}^{\,2}}    \right)^{2} \,\, ,  \quad
 \Lambda_{\pi}= 1700 \, \mbox{MeV} \,\, , 
\end{equation}
 is taken consistently 
from the Elster potential (Table I in \cite{ScA99}). Then,
 we obtain in the c.m.\ frame of the final state the following
  expression for  $j_{6}^{N[1]\, \mu}(z,\vec{k})$:
\begin{eqnarray}
\bra{ {\bar N}(\vec{p}^{\,\prime})}
 j_{6}^{N[1]\, 0}(z,\vec{k})
 \ket{ {\bar N}(\vec{p}\,)} &=& 0 \,, \label{anh1_6}\\
\bra{ {\bar N}(\vec{p}^{\,\prime})}
 \vec{\jmath}_{6}^{\,N[1]}(z,\vec{k})
 \ket{ {\bar N}(\vec{p}\,)} &=&  
 \delta(\vec{p} + \vec{k} - \vec{p}^{\, \prime})\,
\frac{e \tau_0 \,  g_{\pi}^2}{2 M^2_N}  \, 
 \int d^3 q \vec{\Upsilon}(\vec{p}^{\,\prime},\vec{q}) \label{anh1_7}
\end{eqnarray}
with ($\omega_{\pi}(x) = \sqrt{m_{\pi}^2 + x^2}$)
\begin{equation}\label{anh1_7b}
\vec{\Upsilon}(\vec{p}^{\,\prime},\vec{q}) := 
\frac{F^2_{\pi}(\vec{q}^{\,2})}{ (2\pi)^3 2 \omega_{\pi}(\vec{q})}
\frac{\vec{\sigma} \cdot \vec{q} \, \vec{\sigma}
 }{z_{sub} -
 M_N -\frac{(\vec{p}^{\,\prime}- \vec{q} )^2}{2 M_N} - \omega_{\pi}(\vec{q})
}\, \, ,
 \end{equation}
 where for the sake of simplicity
 the argument $i$ in  $j_{6}^{N[1]\, \mu}(z,\vec{k},i)$
 denoting the ``active'' nucleon
 (see (\ref{app_1})) has been skipped.

Note that we use the nonrelativistic energy-momentum relation for the
 ``active'' nucleon; $z_{sub}$ is given  by (\ref{5_zsub}).
 Concerning the  dependence of 
 $\vec{\Upsilon}(\vec{p}^{\,\prime},\vec{q})$ on the angle
 $x :=  \hat{p}^{\,\prime} \cdot \hat{q}$, we make use of the
 Legendre polynomial expansion
\begin{equation}\label{anh1_8}
 \vec{\Upsilon}(\vec{p}^{\,\prime},\vec{q}) \equiv
\vec{\Upsilon}(p',q,x)
= \sum_{\Lambda=0}^{\infty} \vec{\Upsilon}_{\Lambda}(p',q) P_{\Lambda}(x)
\end{equation}
with
\begin{equation}\label{1_9}
\vec{\Upsilon}_{\Lambda}(p',q)
 = \frac{2 \Lambda+1}{2} \int_{-1}^{1} dy \vec{\Upsilon}(p',q,y)
 P_{\Lambda}(y) \,\, .
\end{equation}
 With the help of
\begin{equation}\label{1_10}
\int d\Omega_q \vec{\sigma} \cdot \hat{q}   P_{\Lambda}(\hat{p}^{\,\prime}
 \cdot \hat{q}) = \frac{4\pi}{3} \vec{\sigma} \cdot \hat{p}^{\,\prime}
 \delta_{\Lambda 1}
\end{equation}
 one obtains after some straightforward algebra
\begin{eqnarray}
\bra{ {\bar N}(\vec{p}^{\,\prime})}
 \vec{\jmath}_{6}^{\,N[1]}(z,\vec{k})
 \ket{ {\bar N}(\vec{p}\,)} &=&  
 \delta(\vec{p} + \vec{k} - \vec{p}^{\, \prime})\,
  \, \frac{ e \tau_0\, g_{\pi}^2}{16 \pi^2 M^2_N}
 \, \vec{\sigma} \cdot \hat{p}^{\,\prime} \vec{\sigma}  \quad 
  \nonumber \\ \times \,
 & &  \int d q \frac{q^3 F^2_{\pi}(\vec{q}^{\,2})}{\omega_{\pi}(q)} 
 \int_{-1}^{1} dy \, \frac{y}{z_{sub} - M_N - \frac{p'^2}{2 M_N} -
 \frac{q^2}{2 M_N} + \frac{p' q y }{M_N}} \, \, . \label{anh1_11}  
\end{eqnarray} 
The integral over $y$ can be evaluated with the help
 of the general formula 
\begin{equation}\label{anh1_12}
\int_{-1}^{1} dx \frac{\phi(x)}{x-x_0 + i \epsilon} =  
\PP \int_{-1}^{1} dx \frac{\phi(x)}{x-x_0 }  + \phi(x_0) \
 ln\left|\frac{1-x_0}{1+x_0}\right|
- i \pi \phi(x_0)  \,\, .
\end{equation}
  Therefore, one obtains both 
 real and imaginary parts for the current contribution.
 Finally, with the help of
\begin{equation}\label{anh1_13}
 \vec{\sigma} \cdot \hat{p}^{\,\prime} \vec{\sigma}
 = \hat{p}^{\,\prime} + i \vec{\sigma} \times \hat{p}^{\,\prime} \,\, ,
\end{equation}
 one obtains the contributions of 
 $\vec{\jmath}_{6}^{\,N[1]}(z,\vec{k})$ to the functions 
 $\hat{\alpha},...,\hat{\epsilon}$.
 Concerning the other loop contributions in 
  Fig.\ \ref{figem5}, we would like to note that in general not only one,
 but two or even three angles between the momenta, i.e.\
  $\hat{k} \cdot \hat{p}^{\,\prime}$,
 $\hat{q} \cdot \hat{p}^{\,\prime}$,
 $\hat{k} \cdot \hat{q}$ occur in  the corresponding propagators,
 yielding therefore much more complicated expressions. Nevertheless, the
 general principle of evaluating these loop contributions is the same as
  for  $\vec{\jmath}_{6}^{\,N[1]}(z,\vec{k})$.

\begin{thebibliography}{888}

\bibitem{Bin60}
A. M. Bincer, Phys. Rev. {\bf 118}, 855 (1960)

\bibitem{Nym70}
E. M. Nyman, Nucl. Phys. A  {\bf 154}, 97 (1970)

\bibitem{Nym71}
E. M. Nyman, Nucl. Phys. A  {\bf 160}, 517 (1971)

\bibitem{Har72}
M. G. Hare,  Ann. Phys. {\bf 74}, 595 (1972)

\bibitem{BoK93}
J. W. Bos and J. H. Koch, Nucl. Phys. A{\bf 563}, 539 (1993)

\bibitem{NaK87}
H. W. Naus and J. H. Koch, Phys. Rev. C {\bf 36}, 2459 (1987)

\bibitem{TiT90}
P. C. Tiemeijer and J. A. Tjon, Phys. Rev. C {\bf 42}, 599 (1990)

\bibitem{BoS92}
 J. W. Bos, S. Scherer and J. H. Koch, Nucl. Phys. A {\bf 547}, 488
 (1992)

\bibitem{DoS95}
H. C. D\"onges, M. Sch\"afer and U. Mosel, Phys. Rev. C {\bf 51}, 950 (1995)  

\bibitem{KoM98}
S. Kondratyuk, G. Martinus and O. Scholten, Phys. Lett. B {\bf 418},
 20 (1998)

\bibitem{Fea00}
H. W. Fearing,  Few. Body Syst. Suppl. {\bf 12} (2000) 263

\bibitem{Sch01}
S. Scherer and H. W. Fearing, 
Nucl.Phys. A{\bf 684} (2001) 499

\bibitem{SoC92}
X. Song, J. P. Cheng and J. S. McCarthy, Z. Phys. A {\bf 341}, 275 (1992)

\bibitem{KoS99}
S. Kondratyuk,  and O. Scholten, nucl-th/9906044v2


\bibitem{ScA99}
M. Schwamb and H. Arenh\"ovel,  nucl-th/9912017, accepted for publication in
     Nucl. Phys. A

\bibitem{ScA00}
M. Schwamb and H. Arenh\"ovel,  nucl-th/0008034, accepted  for publication in
     Nucl. Phys. A

\bibitem{ScP80}
D. Sch\"utte  and J. da Providencia, Nucl. Phys. A {\bf 338}, 463 (1980)

\bibitem{ElF88}  
C. Elster, W. Ferchl\"ander, K. Holinde, D. Sch\"utte, and R. Machleidt,
 Phys. Rev. C {\bf 37}, 1647 (1988) 

\bibitem{CaM82}
A. Cambi, B. Mosconi, and P. Ricci, Phys. Rev. Lett. {\bf 48}, 462 (1982).

\bibitem{WiL88}
P. Wilhelm, W. Leidemann, and H. Arenh\"ovel, Few-Body Sys. {\bf 3}, 111 (1988).

\bibitem{CrA96}
R. Crawford {\it et al.}, Nucl. Phys. A {\bf 603}, 303 (1996)

\bibitem{ArG84}
J. Arends {\it et al.}, Nucl. Phys. A {\bf 412}, 509 (1984)

\bibitem{BlB95}
G. Blanpied {\it et al.}, Phys. Rev. C {\bf 52}, R455 (1995);
 Phys. Rev. C {\bf 51}, 024604 (2000)


\bibitem{AdB91}
F. Adamian {\it et al.}, J. Phys. G {\bf 17}, 1189 (1991)

\bibitem{IkA79}
H. Ikeda  {\it et al.},  Phys. Rev. Lett. {\bf 42}, 1321 (1979)

\bibitem{BrD80}
A. S. Bratashevskii  {\it et al.},  Sov. J. Nucl. Phys. {\bf 32}, 216 (1980)

\end{thebibliography}
\end {document}